\documentclass[10pt,british,english,american]{article}
\usepackage[T1]{fontenc}
\usepackage[latin9]{inputenc}
\usepackage{listings}
\usepackage[a4paper]{geometry}
\usepackage{fancyhdr}
\usepackage{babel}
\usepackage{prettyref}
\usepackage{float}
\usepackage{amsmath}
\usepackage{amssymb}
\usepackage{graphicx}
\usepackage[authoryear]{natbib}
\usepackage[unicode=true,
 bookmarks=true,bookmarksnumbered=false,bookmarksopen=false,
 breaklinks=false,pdfborder={0 0 1},backref=false,colorlinks=false]
 {hyperref}
\hypersetup{pdftitle={Unification of correlations},
 pdfauthor={Grytskyy, Tetzlaff, Diesmann, Helias},
 pdfkeywords={leaky integrate-and-fire neuron, Hawkes process, correlations, binary neurons, recurrent networks, Ornstein-Uhlenbeck process},
 colorlinks=true,linkcolor=blue,citecolor=blue,urlcolor=blue,filecolor=blue}
\usepackage{breakurl}

\makeatletter

\floatstyle{ruled}
\newfloat{algorithm}{tbp}{loa}
\providecommand{\algorithmname}{Algorithm}
\floatname{algorithm}{\protect\algorithmname}

\newcommand{\lyxaddress}[1]{
\par {\raggedright #1
\vspace{1.4em}
\noindent\par}
}

\newrefformat{cap}{\hyperref[#1]{Figure~\ref{#1}}}
\newrefformat{fig}{\hyperref[#1]{Figure~\ref{#1}}}
\newrefformat{tab}{\hyperref[#1]{Table~\ref{#1}}}
\newrefformat{sec}{\hyperref[#1]{Sec.~\ref{#1}}}
\newrefformat{sub}{\hyperref[#1]{Sec.~\ref{#1}}}
\newrefformat{alg}{\hyperref[#1]{Alg.~\ref{#1}}}
\newrefformat{app}{\hyperref[#1]{App.~\ref{#1}}}

\usepackage[OT1]{fontenc}

\renewenvironment{abstract}
{\noindent{\fontsize{12}{12}\selectfont\textbf{Abstract}}%
\par\vspace{0.5\baselineskip}\noindent}
{\par}

\renewcommand{\@seccntformat}[1]{%
{\csname the#1\endcsname}.\hspace{0.5em}}

\renewcommand{\section}{\@startsection
{section}%
{1}%
{0mm}%
{-\baselineskip}%
{0.5\baselineskip}%
{\fontsize{12}{12}\selectfont\bfseries}}

\renewcommand{\subsection}[1]{\ssubsection{#1.}}
\newcommand{\ssubsection}{\@startsection
{subsection}%
{2}%
{1em}%
{-\baselineskip}%
{-\fontdimen2\font plus -\fontdimen3\font minus -\fontdimen4\font}%
{\fontsize{12}{12}\selectfont}}

\usepackage{graphics}

\renewcommand{\@makecaption}[2]{%
{\parbox[t]{\linewidth}{%
\textbf{#1:} #2
}}}

\makeatother

\addto\captionsbritish{\renewcommand{\algorithmname}{Algorithm}}
\addto\captionsenglish{\renewcommand{\algorithmname}{Algorithm}}

\begin{document}
\let\oldnameref\nameref \renewcommand{\nameref}[1]{``\oldnameref{#1}''}

\global\long\def\ms{\:\mathrm{ms}}
\global\long\def\mV{\:\mathrm{mV}}
\global\long\def\Hz{\:\mathrm{Hz}}
\global\long\def\pA{\:\mathrm{pA}}
\global\long\def\pF{\:\mathrm{pF}}
\global\long\def\mus{\:\mu\mathrm{s}}
\global\long\def\Vth{V_{\theta}}
\global\long\def\Vr{V_{r}}
\global\long\def\PV{P(V)}
\global\long\def\tilPV{\tilde{P}(V)}
\global\long\def\Lfp{L_{\mathrm{FP}}}
\global\long\def\M{\mathbf{M}}
\global\long\def\I{\mathbf{I}}
\global\long\def\Var{\mathrm{Var}}
\global\long\def\E{\mathrm{E}}
\global\long\def\N{\mathbf{N}}
\global\long\def\f{\mathbf{f}}
\global\long\def\h{\mathbf{h}}
\global\long\def\H{\mathbf{H}}
\global\long\def\Q{\mathbf{Q}}
\global\long\def\q{\mathbf{q}}
\global\long\def\D{\mathbf{D}}
\global\long\def\x{\mathbf{x}}
\global\long\def\y{\mathbf{y}}
\global\long\def\T{\mathbf{T}}
\global\long\def\U{\mathbf{U}}
\global\long\def\r{\mathbf{r}}
\global\long\def\c{\mathbf{c}}
\global\long\def\res{\mathrm{Res}}
\global\long\def\ftr{\mathcal{F}}
\global\long\def\n{\mathbf{n}}
\global\long\def\m{\mathbf{m}}
\global\long\def\W{\mathbf{W}}
\global\long\def\X{\mathbf{X}}
\global\long\def\R{\mathbf{R}}
\global\long\def\a{\mathbf{a}}
\global\long\def\w{\mathbf{w}}
\global\long\def\j{\mathbf{j}}
\global\long\def\J{\mathbf{J}}
\global\long\def\C{\mathbf{C}}
\global\long\def\nuo{\nu_{0}}
\global\long\def\taum{\tau_{m}}
\global\long\def\taus{\tau_{s}}
\global\long\def\taur{\tau_{r}}
\global\long\def\nr{n_{r}}
\global\long\def\diag{\mathrm{diag}}
\global\long\def\Em{\mathbf{1}}
\global\long\def\Y{\mathbf{Y}}
\global\long\def\b{\mathbf{b}}
\global\long\def\A{\mathbf{A}}
\global\long\def\In{\mathcal{I}}
\global\long\def\Ex{\mathcal{E}}
\global\long\def\OO{\mathbf{O}}
\global\long\def\s{\mathbf{s}}
\pdfbookmark[1]{Title}{TitlePage}

\title{A unified view on weakly correlated recurrent networks}

\maketitle
\begin{flushleft}

\par\end{flushleft}

\noindent \begin{flushleft}
Dmytro Grytskyy$^{1,\ast}$, Tom Tetzlaff$^{1}$, Markus Diesmann$^{1,2}$,
Moritz Helias$^{1}$
\par\end{flushleft}

\medskip{}

\noindent \begin{flushleft}
\textbf{{1} Inst. of Neuroscience and Medicine (INM-6) and Inst.
for Advanced Simulation (IAS-6)}\\
\textbf{Jülich Research Centre and JARA, Jülich, Germany}\\
\textbf{{2} Medical Faculty}\\
\textbf{RWTH Aachen University, Germany}\\
\medskip{}
\textbf{$\ast$ email: d.grytskyy@fz-juelich.de}
\par\end{flushleft}
\begin{abstract}
The diversity of neuron models used in contemporary theoretical neuroscience
to investigate specific properties of covariances in the spiking activity
raises the question how these models relate to each other. In particular
it is hard to distinguish between generic properties of covariances
and peculiarities due to the abstracted model. Here we present a unified
view on pairwise covariances in recurrent networks in the irregular
regime. We consider the binary neuron model, the leaky integrate-and-fire
model, and the Hawkes process. We show that linear approximation maps
each of these models to either of two classes of linear rate models,
including the Ornstein-Uhlenbeck process as a special case. The distinction
between both classes is the location of additive noise in the rate
dynamics, which is located on the output side for spiking models and
on the input side for the binary model. Both classes allow closed
form solutions for the covariance. For output noise it separates into
an echo term and a term due to correlated input. The unified framework
enables us to transfer results between models. For example, we generalize
the binary model and the Hawkes process to the situation with synaptic
conduction delays and simplify derivations for established results.
Our approach is applicable to general network structures and suitable
for the calculation of population averages. The derived averages are
exact for fixed out-degree network architectures and approximate for
fixed in-degree. We demonstrate how taking into account fluctuations
in the linearization procedure increases the accuracy of the effective
theory and we explain the class dependent differences between covariances
in the time and the frequency domain. Finally we show that the oscillatory
instability emerging in networks of integrate-and-fire models with
delayed inhibitory feedback is a model-invariant feature: the same
structure of poles in the complex frequency plane determines the population
power spectra.
\end{abstract}

\lyxaddress{}

\section*{Keywords}

Correlations, linear response, Hawkes process, leaky integrate-and-fire
model, binary neuron, linear rate model, Ornstein-Uhlenbeck process

\section{Introduction}

The meaning of correlated neural activity for the processing and representation
of information in cortical networks is still not understood, but evidence
for a pivotal role of correlations increases \citep[recently reviewed in ][]{Cohen11_811}.
Different studies have shown that correlations can either decrease
\citep{Zohary94_140} or increase \citep{Sompolinsky01a} the signal
to noise ratio of population signals, depending on the readout mechanism.
The architecture of cortical networks is dominated by convergent and
divergent connections among the neurons \citep{Braitenberg91} causing
correlated neuronal activity by common input from shared afferent
neurons in addition to direct connections between pairs of neurons
and common external signals. It has been shown that correlated activity
can faithfully propagate through convergent-divergent feed forward
structures, such as synfire chains \citep{Abeles91,Diesmann99_529},
a potential mechanism to convey signals in the brain. Correlated firing
was also proposed as a key to the solution of the binding problem
\citep{Malsburg81,Bienenstock95,Singer99_49}, an idea that has been
discussed controversially \citep{Shadlen99}. Independent of a direct
functional role of correlations in cortical processing, the covariance
function between the spiking activity of a pair of neurons contains
the information about time intervals between spikes. Changes of synaptic
coupling, mediated by spike-timing dependent synaptic plasticity \citep[STDP, ][]{Markram97a,Bi99},
are hence sensitive to correlations. Understanding covariances in
spiking networks is thus a prerequisite to investigate the evolution
of synapses in plastic networks \citep{Burkitt07_533,Gilson09_1,Gilson10_si}.

On the other side, there is ubiquitous experimental evidence of correlated
spike events in biological neural networks, going back to early reports
on multi-unit recordings in cat auditory cortex \citep{Perkel67b,Gerstein69_828},
the observation of closely time-locked spikes appearing at behaviorally
relevant points in time \citep{Kilavik09_12653,Ito11_2482} and collective
oscillations in cortex \citep[recently reviewed in ][]{Buzsaki12_203}. 

The existing theories explaining correlated activity use a multitude
of different neuron models. \citet{Hawkes71_438} developed the theory
of covariances for linear spiking Poisson neurons (Hawkes processes).
\citet{Ginzburg94} presented the approach of linearization to treat
fluctuations around the point of stationary activity and to obtain
the covariances for networks of non-linear binary neurons. The formal
concept of linearization allowed \citet{Brunel99} and \citet{Brunel00_183}
to explain fast collective gamma oscillations in networks of spiking
leaky integrate-and-fire (LIF) neurons. Correlations in feed-forward
networks of leaky integrate-and-fire models are studied in \citet{Morenobote06_028101},
exact analytical solutions for such network architectures are given
in \citet{Rosenbaum10_00116} for the case of stochastic random walk
models, and threshold crossing neuron models are considered in \citet{Tchumatchenko10_058102}
and \citet{Burak09_2269}. Covariances in structured networks are
investigated for Hawkes processes \citep{Pernice11_e1002059}, and
in linear approximation for LIF \citep{Pernice12_031916} and exponential
integrate-and-fire neurons \citep{Trousdale12_e1002408}. The latter
three works employ an expansion of the propagator (time evolution
operator) in terms of the order of interaction. Finally \citet{Buice09_377}
investigate higher order cumulants of the joint activity in networks
of binary model neurons.

\begin{figure}
\begin{centering}
\includegraphics[scale=0.5]{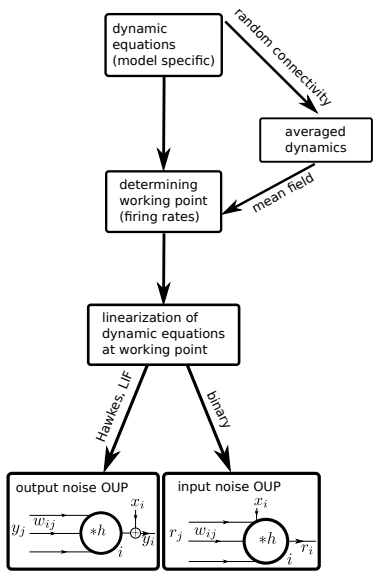}
\par\end{centering}

\caption{Mapping different descriptions of neuronal dynamics to linear rate
models (LRM). The arrows indicate analytical methods which enable
a mapping from the original spiking (leaky integrate-and-fire model,
Hawkes model) or binary neuron dynamics to the analytically more tractable
linear rate models. Depending on the original dynamics (spiking or
binary) the resulting LRM contains an additive noise component $x$
either on the output side (left) or on the input side (right).\foreignlanguage{british}{\label{fig:levels_of_description}}}
\end{figure}

Analytical insight into a neuroscientific phenomenon based on correlated
neuronal activity often requires a careful choice of the neuron model
to arrive at a solvable problem. Hence a diversity of models has been
proposed and is in use. This raises the question which features of
covariances are generic properties of recurrent networks and which
are specific to a certain model. Only if this question can be answered
one can be sure that a particular result is not an artifact of oversimplified
neuronal dynamics. Currently it is unclear how different neuron models
relate to each other and whether and how results obtained with one
model carry over to another. In this work we present a unified theoretical
view on pairwise correlations in recurrent networks in the asynchronous
and collective-oscillatory regime, approximating the response of different
models to linear order. The joint treatment allows us to answer the
question of genericness and moreover naturally leads to a classification
of the considered models into only two categories, as illustrated
in \prettyref{fig:levels_of_description}. The classification in addition
enables us to extend existing theoretical results to biologically
relevant parameters, such as synaptic delays and the presence of inhibition,
and to derive explicit expressions for the time-dependent covariance
functions, in quantitative agreement with direct simulations, which
can serve as a starting point for further work.

The remainder of this article is organized as follows. In the first
part of our results in \nameref{sec:ornstein_uhlenbeck_process} we
investigate the activity and the structure of covariance functions
for two versions of linear rate models (LRM); one with input the other
with output noise. If the activity relaxes exponentially after application
of a short perturbation, both models coincide with the Ornstein-Uhlenbeck
process (OUP). We mainly consider the latter case, although most results
hold for arbitrary kernel functions. We extend the analytical solutions
for the covariances in networks of OUP \citep{Risken96} to the neuroscientifically
important case of synaptic conduction delays. Solutions are derived
first for general forms of connectivity in \nameref{sub:solution_convolution_equation_input_noise}
for input noise and in \nameref{sub:solution_convolution_equation_output_noise}
for output noise. After analyzing the spectral properties of the dynamics
in the frequency domain in \nameref{sub:spectrum_of_dynamics}, identifying
poles of the propagators and their relation to collective oscillations
in neuronal networks, we show in \nameref{sub:population_averaged_correlations}
how to obtain pairwise averaged covariances in homogeneous Erd\"{o}s-R\'{e}nyi
random networks. We explain in detail the use of the residue theorem
to perform the Fourier back-transformation of covariance functions
to the time domain in \nameref{sub:fourier_back_transformation} for
general connectivity and in \nameref{sub:explicit_expressions} for
averaged covariance functions in random networks, which allows us
to obtain explicit results and to discuss class dependent features
of covariance functions.

In the second part of our results in \nameref{sec:binary_neurons},
\nameref{sec:equivalence_hawkes_oup},\\
and \nameref{sec:equivalence_lif_oup} we consider the mapping of
different neuronal dynamics on either of the two flavors of the linear
rate models discussed in the first part. The mapping procedure is
qualitatively the same for all dynamics as illustrated in \prettyref{fig:levels_of_description}:
Starting from the dynamic equations of the respective model, we first
determine the working point described in terms of the mean activity
in the network. For unstructured homogeneous random networks this
amounts to a mean-field description in terms of the population averaged
activity (i.e. firing rate in spiking models). In the next step, a
linearization of the dynamical equations is performed around this
working point. We explain how fluctuations can be considered in the
linearization procedure to improve its accuracy and we show how the
effective linear dynamics maps to the LRM. We illustrate the results
throughout by a quantitative comparison of the analytical results
to direct numerical simulations of the original non-linear dynamics.
The appendices \nameref{sub:implementation_noisy_rate},\\
\nameref{sub:implementation_binary}, and\\
\nameref{sub:implementation_hawkes} describe the model implementations
and are modules of our long-term collaborative project to provide
the technology for neural systems simulations \citep{Gewaltig_07_11204}.

\section{Covariance structure of noisy rate models\label{sec:ornstein_uhlenbeck_process}}

\subsection{Definition of models\label{sub:def_models}}

Let us consider a network of linear model neurons, each characterized
by a continuous fluctuating rate $r$ and connections from neuron
$j$ to neuron $i$ given by the element $w_{ij}$ of the connectivity
matrix $\w$. We assume that the response of neuron $i$ to input
can be described by a linear kernel $h$ so that the activity in
the network fulfills 
\begin{equation}
\r(t)=h(\circ)\ast[\w\r(\circ-d)+\boldsymbol{b}\x(\circ)](t),\label{eq:rdwithb}
\end{equation}
where $f(\circ-d)$ denotes the function $f$ shifted by the delay
$d$, $\x$ is an uncorrelated noise with 

\begin{eqnarray}
\langle x_{i}(t)\rangle & = & 0\text{,\qquad}\langle x_{i}(s)x_{j}(t)\rangle=\delta_{ij}\delta(s-t)\rho^{2}\ ,\label{eq:noise}
\end{eqnarray}
e. g. a Gaussian white noise and $(f\ast g)(t)=\int_{-\infty}^{t}f(t-t^{\prime})g(t^{\prime})\, dt^{\prime}$
is the convolution. With the particular choice $\boldsymbol{b}=\w\delta(\circ-d)\ast$
we obtain

\begin{equation}
\r(t)=[h(\circ)\ast\w(\r(\circ-d)+\x(\circ-d))](t).\label{eq:rdon}
\end{equation}
We call the dynamics \eqref{eq:rdon} the linear noisy rate model
(LRM) with noise applied to output, as the sum $r+x$ appears on the
right hand side. Alternatively, choosing $\boldsymbol{b}=\Em$ we
define the model with input noise as

\begin{equation}
\r(t)=h(\circ)\ast[\w\r(\circ-d)+\x(\circ)](t).\label{eq:rdin}
\end{equation}
Hence, equations \eqref{eq:rdon} and \eqref{eq:rdin} are special
cases of \eqref{eq:rdwithb}. In the following we consider the particular
case of an exponential kernel 
\begin{align}
h(s) & =\frac{1}{\tau}\theta(s)\, e^{-s/\tau},\label{eq:exp_kernel-1}
\end{align}
where $\theta$ denotes the Heaviside function, $\theta(t)=1$ for
$t>0$, $0$ else. Applying to \eqref{eq:rdwithb} the operator $O=\tau\frac{d}{ds}+1$
which has $h$ as a Green's function (i.e. $Oh=\delta)$ we get 

\begin{equation}
\tau\frac{d}{dt}\r(t)+\r(t)=\w\r(t-d)+\b\x(t),\label{eq:OUPin-1}
\end{equation}
which is the equation describing a set of delay coupled Ornstein-Uhlenbeck-processes
(OUP) with input or output noise for $\boldsymbol{b}=\Em$ or $\boldsymbol{b}=\w\delta(\circ-d)\ast$,
respectively. We use this representation in \nameref{sec:binary_neurons}
to show the correspondence to networks of binary neurons.

\subsection{Solution of the convolution equation with input noise\label{sub:solution_convolution_equation_input_noise}}

The solution for the system with input noise obtained from the definition
\eqref{eq:rdin} after Fourier transformation is

\begin{equation}
\R=H_{d}\w\R+H\X,
\end{equation}
where the delay is consumed by the kernel function $h_{d}(s)=\frac{1}{\tau}\theta(s-d)e^{-(s-d)/\tau}$.
We use capital letters throughout the text to denote objects in the
Fourier domain and lower case letters for objects in the time domain.
Solved for $\R=(1-H_{d}\w)^{-1}H\X$ the covariance function of $\r$
in the Fourier domain is found with the Wiener\textendash{}Khinchin
theorem \citep{Gardiner04} as $\langle\R(\omega)\R^{T}(-\omega)\rangle$,
also called the cross spectrum

\begin{align}
\C(\omega) & =\langle\R(\omega)\R^{T}(-\omega)\rangle\label{eq:cross_spectrum_in}\\
 & =(1-H_{d}(\omega)\w)^{-1}H(\omega)\langle\X(\omega)\X^{T}(-\omega)\rangle H(-\omega)(1-H_{d}(-\omega)\w^{T})^{-1}\nonumber \\
 & =(H_{d}(\omega)^{-1}-\w)^{-1}\D(H_{d}(-\omega)^{-1}-\w^{T})^{-1},\nonumber 
\end{align}
where we introduced the matrix $\D=\langle\X(\omega)\X^{T}(-\omega)\rangle$.
From the second to the third line we used the fact that the non-delayed
kernels $H(\omega)$ can be replaced by delayed kernels $H_{d}(\omega)$
and that the corresponding phase factors $e^{i\omega d}$ and $e^{-i\omega d}$
cancel each other. If $\x$ is a vector of pairwise uncorrelated noise,
$\D$ is a diagonal matrix and needs to be chosen accordingly in order
for the cross spectrum \eqref{eq:cross_spectrum_in} to coincide (neglecting
non-linear effects) with the cross spectrum of a network of binary
neurons, as described in\\
\nameref{sub:equivalence_binary_oup}.

\subsection{Solution of convolution equation with output noise\label{sub:solution_convolution_equation_output_noise}}

For the system with output noise we consider the quantity $y_{i}=r_{i}+x_{i}$
as the dynamic variable representing the activity of neuron $i$ and
aim to determine pairwise correlations. It is easy to get from \eqref{eq:rdon}
after Fourier transformation

\begin{equation}
\R=H_{d}\w(\R+\X),
\end{equation}
which can be solved for $\R=(1-H_{d}\w)^{-1}H_{d}\w\X$ in order to
determine the Fourier transform of $\Y$ as

\begin{equation}
\Y=\R+\X=(1-H_{d}\w)^{-1}\X.
\end{equation}
The cross spectrum hence follows as

\begin{align}
\C(\omega) & =\langle\Y(\omega)\Y{}^{T}(-\omega)\rangle\label{eq:cross_spectrum_on}\\
 & =(1-H_{d}(\omega)\w)^{-1}\langle\X(\omega)\X^{T}(-\omega)\rangle(1-H_{d}(-\omega)\w^{T})^{-1}\nonumber \\
 & =(1-H_{d}(\omega)\w)^{-1}\D(1-H_{d}(-\omega)\w^{T})^{-1},\nonumber 
\end{align}
with $\D=\langle\X(\omega)\X^{T}(-\omega)\rangle$. $\D$ is a diagonal
matrix with the $i$-th diagonal entry $\rho_{i}^{2}$. For the correspondence
to spiking models $\D$ must be chosen appropriately, as discussed
in \nameref{sec:equivalence_hawkes_oup} and \nameref{sec:equivalence_lif_oup}
for Hawkes processes and leaky integrate-and-fire neurons, respectively.

\subsection{Spectrum of the dynamics\label{sub:spectrum_of_dynamics}}

For both linear rate dynamics, with output and with input noise, the
cross spectrum $\C(\omega)$ has poles at certain frequencies $\omega$
in the complex plane. These poles are defined by the zeros of $\det(H_{d}(\omega)^{-1}-\w)$
and the corresponding term with the opposite sign of $\omega$. The
zeros of $\det(H_{d}(\omega)^{-1}-\w)$ are solutions of the equation 

\[
H_{d}(\omega)^{-1}=(1+i\omega\tau)e^{i\omega d}=L_{j}
\]
where $L_{j}$ is the $j$-th eigenvalue of $\w$. The same set of
poles arises from \eqref{eq:rdwithb} when solving for $\R$. For
$d>0$ and the exponential kernel \eqref{eq:exp_kernel-1}, the poles
can be expressed as

\begin{equation}
z_{k}(L_{j})=\frac{i}{\tau}-\frac{i}{d}W_{k}(L_{j}\frac{d}{\tau}e^{\frac{d}{\tau}}),\label{eq:poles_z}
\end{equation}
where $W_{k}$ is the $k$-th of the infinitely many branches of the
Lambert-W function \citep{Corless96_329}. For vanishing synaptic
delay $d=0$ there is obviously only one solution for every $L_{j}$
given by $z=\frac{-i}{\tau}(L_{j}-1)$.

Given the same parameters $d$, $\w$, $\tau$, the pole structures
of the cross spectra of both systems \eqref{eq:cross_spectrum_in}
and \eqref{eq:cross_spectrum_on} are identical, since the former
can be obtained from the latter by multiplication with $(H_{d}(\omega)H_{d}(-\omega))^{-1}=(H(\omega)H(-\omega))^{-1}$,
which has no poles. The only exception causing a different pole structure
for the two models is the existence of an eigenvalue $L_{j}=0$ of
the connectivity matrix $\w$, corresponding to a pole $z(0)=\frac{i}{\tau}$.
However, this pole corresponds to an exponential decay of the covariance
for input noise in the time domain and hence does not contribute to
oscillations. For output noise, the multiplication with the term \foreignlanguage{english}{\textrm{$(H(\omega)H(-\omega))^{-1}$}}\textrm{,}
vanishing at $\omega=\frac{i}{\tau}$, cancels this pole in the covariance.
 Consequently both dynamics exhibit similar oscillations. A typical
spectrum of poles for a negative eigenvalue $L_{j}<0$ is shown in
\prettyref{fig:power_spectra}B,D.

\subsection{Population-averaged covariances\label{sub:population_averaged_correlations}}

Often it is desirable to consider not the whole covariance matrix
but averages over subpopulations of pairs of neurons. For instance
the average over the whole network would result in a single scalar
value. Separately averaging pairs, distinguishing excitatory and inhibitory
neuron populations, yields a $2$ by $2$ matrix of covariances. For
these simpler objects closed form solutions can be obtained, which
already preserve some useful information and show important features
of the network. Averaged covariances are also useful for comparison
with simulations and experimental results.

In the following we consider a recurrent random network of $N_{e}=N$
excitatory and $N_{i}=\gamma N$ inhibitory neurons with synaptic
weight $w$ for excitatory and $-gw$ for inhibitory synapses. The
probability $p$ determines the existence of a connection between
two randomly chosen neurons. We study the dynamics averaged over the
two subpopulations by introducing the quantities $r_{a}=\frac{1}{N_{a}}\sum_{j\in a}r_{j}$
and noise terms $x_{a}=\frac{1}{N_{a}}\sum_{j\in a}x_{j}$ for $a\in\{\Ex,\In\}$;
indices $\In$ and $\Ex$ stand for inhibitory and excitatory neurons
and corresponding quantities. Calculating the average local input
$N_{a}^{-1}\sum_{j\in a}w_{jk}r_{k}$ to a neuron of type $a$, we
obtain

\begin{align}
N_{a}^{-1}\sum_{j\in a}\sum_{k}w_{jk}r_{k} & =\label{eq:average_connectivity}\\
 & =N_{a}^{-1}(\sum_{j\in a}\sum_{k\in\Ex}w_{jk}r_{k}+\sum_{j\in a}\sum_{k\in\In}w_{jk}r_{k})\nonumber \\
 & =N_{a}^{-1}(pN_{a}w\sum_{k\in\Ex}r{}_{k}-pN_{a}gw\sum_{k\in\In}r{}_{k})\nonumber \\
 & =pwN(r_{\Ex}-\gamma gr_{\In}),\nonumber 
\end{align}
where, from the second to the third line we used the fact that in
expectation a given neuron $k$ has $pN_{a}$ targets in the population
$a$. The reduction to the averaged system in \eqref{eq:average_connectivity}
is exact if in every column $k$ in $\w_{jk}$ there are exactly $K$
non-zero elements for $j\in\Ex$ and $\gamma K$ for $j\in\In$, which
is the case for networks with fixed out-degree (number of outgoing
connections of a neuron to the neurons of a particular type is kept
constant), as noted earlier \citep{Tetzlaff12_e1002596}. For fixed
in-degree (number of connections to a neuron coming in from the neurons
of a particular type is kept constant) the substitution of $r_{j\in a}$
by $r_{a}$ is an additional approximation, which could be considered
as an average over possible realizations of the random connectivity.
In both cases the effective population-averaged connectivity matrix
$\M$ turns out to be

\begin{equation}
\M=Kw\left(\begin{array}{cc}
1 & -\gamma g\\
1 & -\gamma g
\end{array}\right),\label{eq:averaged_conn}
\end{equation}
with $K=pN$. So the averaged activities fulfill the same equations
\eqref{eq:rdon} and \eqref{eq:rdin} with the non-averaged quantities
$\r$, $\x$, and $\w$ replaced by their averaged counterparts $\bar{\r}=(r_{\Ex},r_{\In})^{T}$,
$\bar{\x}=(x_{\Ex},x_{\In})^{T}$, and $\M$. The population averaged
activities $r_{a}$ are directly related to the block-wise averaged
covariance matrix $\bar{\c}=\left(\begin{array}{cc}
c_{\Ex\Ex} & c_{\Ex\In}\\
c_{\In\Ex} & c_{\In\In}
\end{array}\right)$, with $c_{ab}=N_{a}^{-1}N_{b}^{-1}\sum_{i\in a}\sum_{j\in b}c_{ij}$.
With
\begin{align}
\bar{D}_{ab}= & N_{a}^{-1}N_{b}^{-1}\left\langle \sum_{i\in a}x_{i}\sum_{j\in b}x_{j}\right\rangle \label{eq:averagedD}\\
= & N_{a}^{-1}N_{b}^{-1}\sum_{i\in a}\sum_{j\in b}D_{ij}\nonumber \\
= & \delta_{ab}N_{a}/N_{a}^{2}\rho^{2}=\delta_{ab}N_{a}^{-1}\rho^{2}\nonumber 
\end{align}
we replace $\D$ by $\bar{\D}=\rho^{2}\left(\begin{array}{cc}
N^{-1} & 0\\
0 & (\gamma N)^{-1}
\end{array}\right)$ and $\c$ by $\bar{\c}$ so that the same equations \eqref{eq:cross_spectrum_on}
and \eqref{eq:cross_spectrum_in} and their general solutions also
hold for the block-wise averaged covariance matrices.

\begin{figure}
\begin{centering}
\includegraphics{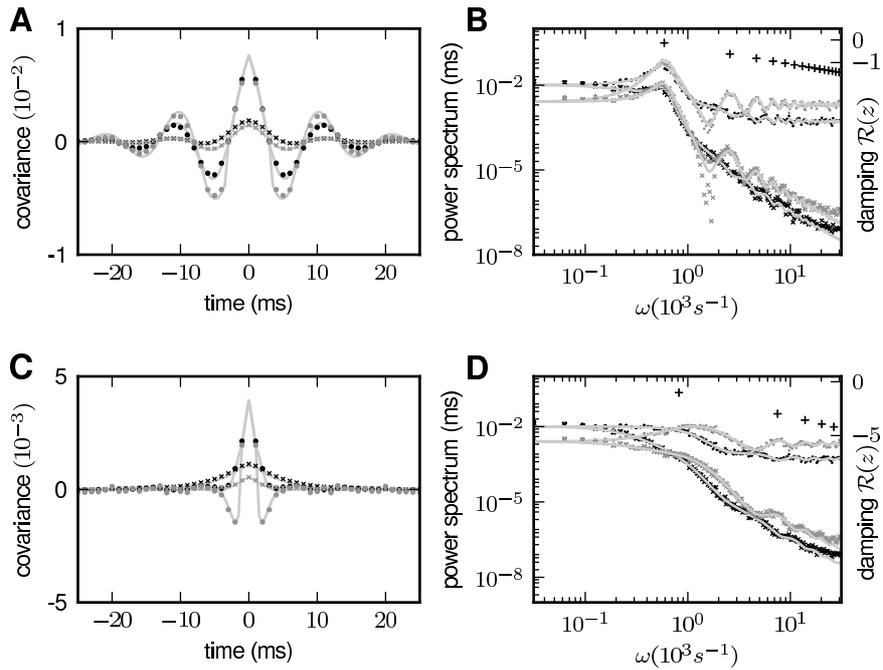}
\par\end{centering}

\caption{Pole structure determines dynamics. Autocovariance of the population
activity (\textbf{A},\textbf{C}) measured in $\rho^{2}/\tau$ and
its Fourier transform called power spectrum (\textbf{B},\textbf{D})
of the rate models with output noise (dots) \eqref{eq:rdon} and input
noise (diagonal crosses) \eqref{eq:rdin} for delays $d=3\ms$ (\textbf{A},\textbf{B})
and $d=1\ms$ (\textbf{C},\textbf{D}). Black symbols show averages
over the excitatory population activity and gray symbols over the
inhibitory activity obtained by direct simulation. Light gray curves
show theoretical predictions for the spectrum \eqref{eq:avg_cross_spectrum_on}
and the covariance \eqref{eq:cov_avg_oupon_t} for output noise and
the spectrum \eqref{eq:avg_cross_spectrum_in} and the covariance
\eqref{eq:cov_avg_oupin_t} for input noise. Black crosses \eqref{eq:poles_z}
in B, D denote the locations of the poles of the cross spectra - with
the real parts corresponding to the damping (vertical axis), and the
imaginary parts to oscillation frequencies (horizontal axis). The
detailed parameters for this and following figures are given in \nameref{sub:parameters}.
\label{fig:power_spectra}}
\end{figure}
The covariance matrices separately averaged over pairs of excitatory,
inhibitory or mixed pairs are shown in \prettyref{fig:power_spectra}
for both linear rate dynamics \eqref{eq:rdon} and \eqref{eq:rdin}.
(Parameters for all simulations presented in this article are collected
in \nameref{sub:parameters}, the implementation of linear rate models
is described in \nameref{sub:implementation_noisy_rate}). The poles
of both models shown in \prettyref{fig:power_spectra}B are given
by \eqref{eq:poles_z} and coincide with the peaks in the cross spectra
\eqref{eq:cross_spectrum_in} and \eqref{eq:cross_spectrum_on} for
output and input noise, respectively. The results of direct simulation
and the theoretical prediction are shown for two different delays,
with the longer delay leading to stronger oscillations.

\begin{figure}
\begin{centering}
\includegraphics{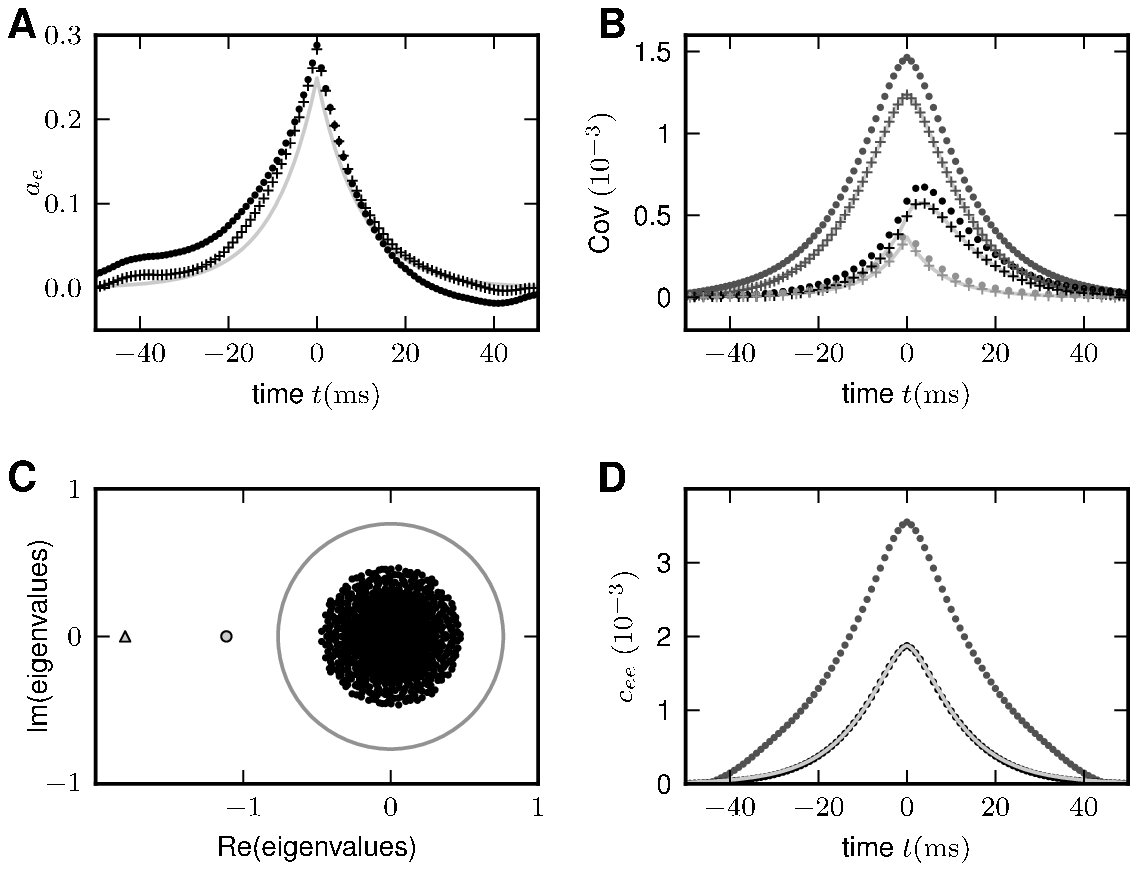}
\par\end{centering}

\caption{Limits of the theory for fixed in-degree and fixed out-degree. Autocovariance
(\textbf{A}) and covariance (\textbf{B}) in random networks with fixed
in-degree (dots) and fixed out-degree (crosses). Simulation results
for $c_{\Ex\Ex}$, $c_{\Ex\In}$, and $c_{\In\In}$ are shown in dark
gray, black and light gray, respectively for synaptic weight $w=0.011$
far from bifurcation. For larger synaptic weight $w=0.018$ close
to bifurcation (see text at the end of \nameref{sub:population_averaged_correlations}),
$c_{\Ex\Ex}$ is also shown in \textbf{D} for fixed in-degree (dark
gray dots) and for fixed out-degree (black dots). Corresponding theoretical
predictions for the autocovariance \eqref{eq:auto_cov_in} (A) and
the covariance \eqref{eq:cov_avg_oupin_t} (B,D) are plotted as light
gray curves throughout. The set of eigenvalues is shown as black dots
in panel \textbf{C} for the smaller weight. The gray circle denotes
the spectral radius $w\sqrt{Np(1-p)(1+\gamma g^{2})}$ \citep{Rajan06,Kriener08_2185}
confining the set of eigenvalues for the larger weight. The small
filled gray circle and the triangle show the effective eigenvalues
$L$ of the averaged systems for small and large weight, respectively.\label{fig:invsoutdegree}}
\end{figure}
\prettyref{fig:invsoutdegree}C shows the distribution of eigenvalues
in the complex plane for two random connectivity matrices with different
synaptic amplitudes $w$. The model exhibits a bifurcation, if at
least one eigenvalue assumes a zero real part. For fixed out-degree
the averaging procedure \eqref{eq:average_connectivity} is exact,
reflected by the precise agreement of theory and simulation in \prettyref{fig:invsoutdegree}D.
For fixed in-degree, the averaging procedure \eqref{eq:average_connectivity}
is an approximation, which is good only for parameters far from the
bifurcation. Even in this regime still small deviations of the theory
from the simulation results are visible in \prettyref{fig:invsoutdegree}B.
On the stable side close to a bifurcation, the appearance of long
living modes causes large fluctuations. These weakly damped modes
appearing in one particular realization of the connectivity matrix
are not represented after the replacement of the full matrix $\w$
by the average $\M$ over matrix realizations. The eigenvalue spectrum
of the connectivity matrix provides an alternative way to understand
the deviations. By the averaging the set of $N$ eigenvalues of the
connectivity matrix is replaced withby the two eigenvalues of the
reduced matrix $\M$, one of which is zero due to identical rows of
$\M$. The eigenvalue spectrum of the full matrix is illustrated in
\prettyref{fig:invsoutdegree}C. Even if the eigenvalue(s) $L^{\M}$
of $\M$ are far in the stable region (corresponding to $\Im(z(L^{\M}))>0$)
some eigenvalues $L^{\w}$ of the full connectivity matrix in the
vicinity of the bifurcation region may still have an imaginary part
becoming negative and the system can feel their influence, shown in
\prettyref{fig:invsoutdegree}D.

\subsection{Fourier back transformation\label{sub:fourier_back_transformation}}

Although the cross spectral matrices \prettyref{eq:cross_spectrum_in}
and \prettyref{eq:cross_spectrum_on} for both dynamics look similar
in the Fourier domain, the procedures for back transformation differ
in detail. In both cases, the Fourier integral along the real $\omega$-axis
can be extended to a closed integration contour by a semi-circle with
infinite radius centered at $0$ in the appropriately chosen half-plane.
The half-plane needs to be selected such that the contribution of
the integration along the semi-circle vanishes. By employing the residue
theorem \citep{Bronstein99} the integral can be replaced by a sum
over residua of the poles encircled by the contour. For a general
covariance matrix we only need to calculate $\c(t)$ for $t\geq0$,
as for $t<0$ the solution can be found by symmetry $\c(-t)=\c^{T}(-t)$.

For input noise it is possible to close the contour in the upper half-plane
where the integrand $\C(\omega)\, e^{i\omega t}$ vanishes for $|\omega|\rightarrow\text{\ensuremath{\infty}}$
for all $t>0$, as $|C_{ij}(\omega)|$ decays as $|\omega|^{-2}$.
This can be seen from \eqref{eq:cross_spectrum_in}, because the highest
order of $H_{d}^{-1}\propto\omega$ appearing in $\det(H_{d}^{-1}-\w)$
is equal to the dimensionality $N$ of $\w$ ($N=2$ for $\M$), and
in $\det(\text{adjugate matrix \ensuremath{ij}of }H_{d}^{-1}-\w)$
it is $N-1$ ($i=j$) or $N-2$ ($i\neq j$). So $|(H_{d}^{-1}-\w)^{-1}|$
is proportional to $|\omega|^{-1}|e^{-i\omega d}|$ and $|\C(\omega)|\propto|\omega|^{-2}$
for large $|\omega|$.

For the case of output noise \eqref{eq:cross_spectrum_on} $\C(\omega)$
can be obtained from the $\C(\omega)$ for input noise \eqref{eq:cross_spectrum_in}
multiplied with $(H_{d}(\omega)H_{d}(-\omega))^{-1}\sim|\omega|^{2}$
for large $|\omega|$. The multiplication with this factor changes
the asymptotic behavior of the integrand, which therefore contains
terms converging to a constant value and terms decaying like $|\omega|^{-1}$
for $|\omega|\rightarrow\text{\ensuremath{\infty}}$. These terms
result in non-vanishing integrals over the semicircle in the upper
half-plane and have to be considered separately. To this end we rewrite
\eqref{eq:cross_spectrum_on} as

\begin{align}
\C(\omega)= & ((1-H_{d}(\omega)\w)^{-1}H_{d}(\omega)\w+1)\D(\w^{T}H_{d}(-\omega)(1-H_{d}(-\omega)\w^{T})^{-1}+1)\label{eq:cross_spectrum_on_separated}\\
= & (1-H_{d}(\omega)\w)^{-1}H_{d}(\omega)\w\D\w^{T}H_{d}(-\omega)(1-H_{d}(-\omega)\w^{T})^{-1}\nonumber \\
+ & (1-H_{d}(\omega)\w)^{-1}H_{d}(\omega)\w\D\nonumber \\
+ & \D\w^{T}H_{d}(-\omega)(1-H_{d}(-\omega)\w^{T})^{-1}\nonumber \\
+ & \D,\nonumber 
\end{align}
and find the constant term $\D$ which turns into a $\delta$-function
in the time domain. The first term in the second line of \eqref{eq:cross_spectrum_on_separated}
decays like $|\omega|^{-2}$ and can be transformed just as $\C(\omega)$
for input noise closing the contour in the upper half-plane. The second
and third term are the transposed complex conjugates of each other,
because of the dependence of $H$ on $-\omega$ instead of $\omega$,
and require a special consideration. Multiplied by $e^{i\omega t}$
under the Fourier integral, the first term is proportional to $H_{d}e^{i\omega t}\sim\omega^{-1}e^{i\omega(t-d)}$
and vanishes faster than $|\omega|^{-1}$ for large $|\omega|$ in
the upper half-plane for $t>d$ and in the lower half plane for $t<d$.
For the second term the half planes are interchanged. The application
of the residue theorem requires closing the integration contour in
the half-plane where the integral over the semi-circle vanishes faster
than $|\omega|^{-1}$. For $\w=\M$ and in the general case of a stable
dynamics all poles of the first term are in the upper half-plane $\Im(z_{k}(L_{j}))>0$,
and have no contribution to $\c(t)$ for $t<d$. For the second term
the same is true for $t>-d$; these terms correspond to the jumps
of $\c(t)$ after one delay, caused by the effect of the sending neuron
arriving at the other neurons in the system after one synaptic delay.
These terms correspond to the response of the system to the impulse
of the sending neuron -- hence we call them {}``echo terms'' in
the following \citep{Helias13_023002}. The presence of such discontinuous
jumps at time points $d$ and $-d$ in the case of output noise is
reflected in the convolution of $h\w$ with $\D$ in the time domain
in \eqref{eq:convolution_eq_oup_on}. For input noise the absence
of discontinuities can be inferred from the absence of such terms
in \eqref{eq:masterforoupin}, where the derivative of the correlation
function is equal to the sum of finite terms. The first summand in
\eqref{eq:cross_spectrum_on_separated} corresponds to the covariance
evoked by fluctuations propagating through the system originating
from the same neuron and we call it {}``correlated input term''.
In the system with input noise a similar separation into effective
echo and correlated input terms can be performed. We obtain the correlated
input term as the covariance in an auxiliary population without outgoing
connections and echo terms as the difference between the full covariance
between neurons within the network and the correlated input term.

\subsection{Explicit expression for the population averaged cross covariance
in the time domain\label{sub:explicit_expressions}}

We obtain the population averaged cross spectrum in a recurrent random
network of Ornstein-Uhlenbeck processes with input noise by inserting
the averaged connectivity matrix $\w=\M$ \eqref{eq:averaged_conn}
into \eqref{eq:cross_spectrum_in}. The explicit expression for the
covariance function follows by taking into account all (both) eigenvalues
of $\M$ with values $0$ and $L=Kw(1-\gamma g)$. The detailed derivation
of the results presented in this section are documented in \nameref{sub:explicit_expressions_app}.
The expression for the cross spectrum \eqref{eq:cross_spectrum_in}
takes the form

\begin{align}
\C(\omega) & =f(\omega)f(-\omega)\left(\Em+Kw\left(\begin{array}{cc}
\gamma g & -\gamma g\\
1 & -1
\end{array}\right)H_{d}(\omega)\right)\D\left(\Em+Kw\left(\begin{array}{cc}
\gamma g & 1\\
-\gamma g & -1
\end{array}\right)H_{d}(-\omega)\right),\label{eq:avg_cross_spectrum_in}
\end{align}
where we introduced $f(\omega)=(H_{d}(\omega)^{-1}-L){}^{-1}$ as
a short hand. Sorting the terms by their dependence on $\omega$,
introducing the functions $\Phi_{1}(\omega),\ldots,\Phi_{4}(\omega)$
for this dependence, and $\varphi_{1}(t),\ldots,\varphi_{4}(t)$ for
the corresponding functions in the time domain, the covariance in
the time domain $\c(t)=\frac{1}{2\pi}\int_{-\infty}^{+\infty}\C(\omega)e^{i\omega t}d\omega$
takes the form

\begin{align*}
\c(t) & =\D\varphi_{1}(t)\\
 & +Kw\left(\left(\begin{array}{cc}
\gamma g & -\gamma g\\
1 & -1
\end{array}\right)\D\varphi_{2}(t)+\D\left(\begin{array}{cc}
\gamma g & 1\\
-\gamma g & -1
\end{array}\right)\varphi_{3}(t)\right)\\
 & +K^{2}w^{2}\left(\begin{array}{cc}
\gamma g & -\gamma g\\
1 & -1
\end{array}\right)\D\left(\begin{array}{cc}
\gamma g & 1\\
-\gamma g & -1
\end{array}\right)\varphi_{4}(t).
\end{align*}
The previous expression is valid for arbitrary $\D$. In simulations
presented in this article we consider identical marginal input statistics
for all neurons. In this case the averaged activities for excitatory
and inhibitory neurons are the same, so we can insert the special
form of $\D$ given in \eqref{eq:averagedD}, which results in 
\begin{align}
\c(t) & =\frac{\rho^{2}}{N}\left(\begin{array}{cc}
1 & 0\\
0 & \gamma^{-1}
\end{array}\right)\varphi_{1}(t)\label{eq:cov_avg_oupin_t}\\
 & +\frac{\rho^{2}}{N}Kw\left(\begin{array}{cc}
\gamma g & -g\\
1 & -\gamma^{-1}
\end{array}\right)\varphi_{2}(t)+\frac{\rho^{2}}{N}Kw\left(\begin{array}{cc}
\gamma g & 1\\
-g & -\gamma^{-1}
\end{array}\right)\varphi_{3}(t)\nonumber \\
 & +\frac{\rho^{2}}{N}(\gamma+1)K^{2}w^{2}\left(\begin{array}{cc}
\gamma g^{2} & g\\
g & \gamma^{-1}
\end{array}\right)\varphi_{4}(t).\nonumber 
\end{align}
The time-dependent functions $\varphi_{1},\ldots,\varphi_{4}$ are
the same in both cases. Using the residue theorem $\varphi_{i}(t)=\frac{1}{2\pi}\int_{-\infty}^{+\infty}\Phi_{i}(\omega)e^{i\omega t}d\omega=i\sum_{z\in\text{poles of }\Phi_{i}}\mathrm{Res}(\Phi_{i},z)\, e^{izt}$
for $t\geqq0$ they can be expressed as a sum over the poles $z_{k}(L)$
given by \eqref{eq:poles_z} and the pole $z=\frac{i}{\tau}$ of $H_{d}(\omega)$.
At $\omega=z_{k}(L)$ the residue of \foreignlanguage{english}{\textrm{$f(\omega)$}}
is $\mathrm{Res}(f,\omega=z_{k}(L))=\left(idL+i\tau e^{i\omega d}\right)^{-1},$
the residue of $H_{d}(\omega)$ at $z=\frac{i}{\tau}$ is $-\frac{i}{\tau}e^{d/\tau}$,
so that the explicit forms of $\varphi_{1},\ldots,\varphi_{4}$ follow
as

\begin{align}
\varphi_{1}(t) & =\sum_{\omega=z_{k}(L)}i\mathrm{Res}(f,\omega)f(-\omega)e^{i\omega t}\nonumber \\
\varphi_{2}(t) & =\sum_{\omega=z_{k}(L)}i\mathrm{Res}(f,\omega)f(-\omega)H_{d}(\omega)e^{i\omega t}+\frac{e^{(d-t)/\tau}}{\tau}f(\frac{i}{\tau})f(-\frac{i}{\tau})\nonumber \\
\varphi_{3}(t) & =\sum_{\omega=z_{k}(L)}i\mathrm{Res}(f,\omega)f(-\omega)H_{d}(-\omega)e^{i\omega t}\label{eq:phis}\\
\varphi_{4}(t) & =\sum_{\omega=z_{k}(L)}i\mathrm{Res}(f,\omega)f(-\omega)H_{d}(\omega)H_{d}(-\omega)e^{i\omega t}+\frac{e^{-t/\tau}}{2\tau}f(\frac{i}{\tau})f(-\frac{i}{\tau}).\nonumber 
\end{align}
The corresponding expression for $\C(\omega)$ for output noise is
obtained by multiplying \eqref{eq:avg_cross_spectrum_in} with $H_{d}^{-1}(\omega)H_{d}^{-1}(-\omega)=(1+\omega^{2}\tau^{2})$

\begin{align}
\C(\omega) & =H_{d}^{-1}(\omega)H_{d}^{-1}(-\omega)f(\omega)f(-\omega)\label{eq:avg_cross_spectrum_on}\\
 & \times(\Em+Kw\left(\begin{array}{cc}
\gamma g & -\gamma g\\
1 & -1
\end{array}\right)H_{d}(\omega))\D(\Em+Kw\left(\begin{array}{cc}
\gamma g & 1\\
-\gamma g & -1
\end{array}\right)H_{d}(-\omega)),\nonumber 
\end{align}
which, after Fourier transform, provides the expression for $\c(t)$
in the time domain for $t\geqq0$

\begin{align}
\c(t) & =\M\D\M^{T}\varphi_{1}(t)+\M\D\varphi_{0}(t)+\D\delta(t)\nonumber \\
 & =K^{2}w^{2}\frac{\rho^{2}}{N}(1+\gamma g^{2})\left(\begin{array}{cc}
1 & 1\\
1 & 1
\end{array}\right)\varphi_{1}(t)+Kw\frac{\rho^{2}}{N}\left(\begin{array}{cc}
1 & -g\\
1 & -g
\end{array}\right)\varphi_{0}(t)+\frac{\rho^{2}}{N}\left(\begin{array}{cc}
1 & 0\\
0 & \gamma{}^{-1}
\end{array}\right)\delta(t).\label{eq:cov_avg_oupon_t}
\end{align}
As in \eqref{eq:cov_avg_oupin_t}, the first line holds for arbitrary
$\D$, and the second for $\D$ given by \eqref{eq:averagedD}, valid
if the firing rates are homogeneous. $\varphi_{1}$ is defined as
before, and

\begin{equation}
\varphi_{0}(t)=\theta(t-d)\sum_{\omega=z_{k}(L)}\left(dL+\tau e^{i\omega d}\right)^{-1}e^{i\omega t}\label{eq:phi0}
\end{equation}
vanishes for $t<d$. All matrix elements of the first term in \eqref{eq:cov_avg_oupon_t}
are identical. Therefore all elements of $\c(t)$ are equal for $0<|t|<d$.
Both rows of the matrix in front of $\varphi_{0}$ are identical,
so for $t>0$ the off diagonal term $c_{\In\Ex}$ coincides with $c_{\Ex\Ex}$
and $c_{\Ex\In}$ with $c_{\In\In}$ and vice versa for $t<0$. 

\begin{figure}
\begin{centering}
\includegraphics{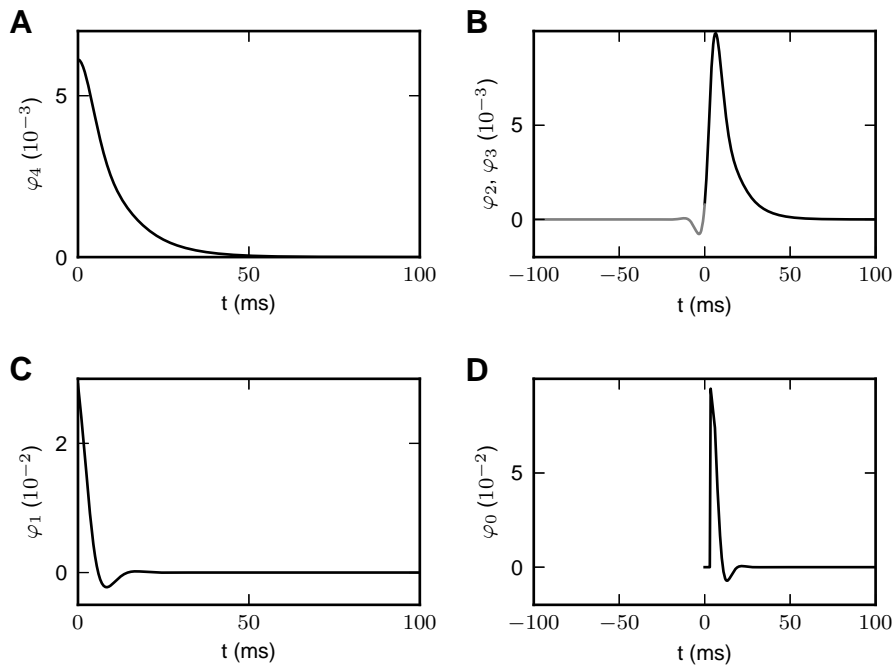}
\par\end{centering}

\caption{Functions $\varphi_{i,\ i=0...4}$ introduced in \eqref{eq:phis}
and \eqref{eq:phi0} for decomposition of covariance $\c(t)$. In
panel \textbf{B} $\varphi_{3}(-t)$ is shown in gray and $\varphi_{2}(t)$
in black. The two functions are continuations of each other, joint
at $t=0$. Both functions appear in the echo term for input noise.
The function $\varphi_{0}$ in panel D describing the corresponding
echo term in the case of output noise is shifted to be aligned with
the function in panel B to facilitate the comparison of panels B and
D. Parameters in all panels are $d=3\ms$, $\tau=10\ms$, $L=-1.72$.\label{fig:phis}}
\end{figure}

As an illustration we show the functions $\varphi_{0},\ldots,\varphi_{4}$
for one set of parameters in \prettyref{fig:phis}. The left panels
(A,C) correspond to contributions to the covariance caused by common
input to a pair of neurons, the right panels (B,D) to terms due to
the effect of one of the neurons' activities on the remaining network
(echo terms). The upper panels (A,B) belong to the model with input
noise, the lower panel (C,D) to the one with output noise.

For the rate dynamics with output noise, the term with $\varphi_{1}$
in \eqref{eq:cov_avg_oupon_t} (shown in \prettyref{fig:phis}C) is
symmetric and describes the common input covariance and the term with
$\varphi_{0}$ (shown in \prettyref{fig:phis}D) is the echo part
of the covariance. For the rate dynamics with input noise \eqref{eq:cov_avg_oupin_t}
the term containing $\varphi_{4}$ (shown in \prettyref{fig:phis}A)
is caused by common input and is hence also symmetric, the terms with
$\varphi_{2}$ and $\varphi_{3}$ (shown in \prettyref{fig:phis}B)
correspond to the echo part and have hence their peak outside the
origin. The second echo term in \eqref{eq:cov_avg_oupin_t} is equal
to the first one transposed and with opposite sign of the time argument,
so we show $\varphi_{2}(t)$ and $\varphi_{3}(-t)$ together in one
panel in \prettyref{fig:phis}B. Note that for input noise, the term
with $\varphi_{1}$ describes the autocovariance, which corresponds
to the term with the $\delta$-function in case of output noise.

The solution \eqref{eq:cov_avg_oupin_t} is visualized in \prettyref{fig:binandlin},
the solution \eqref{eq:cov_avg_oupon_t} in \prettyref{fig:covariance_spiking},
and the decomposition into common input and echo parts is also shown
and compared to direct simulations in \prettyref{fig:echoes}. 

\section{Binary neurons\label{sec:binary_neurons}}

In the following sections we study, in turn, the binary neuron model,
the Hawkes model and the leaky integrate-and-fire model and show how
they can be mapped to one of the two OUPs; either the one with input
or the one with output noise, so that the explicit solutions \eqref{eq:cov_avg_oupin_t}
and \eqref{eq:cov_avg_oupon_t} for the covariances derived in the
previous section can be applied. In the present section, we start
with the binary neuron model \citep{Ginzburg94,Buice09_377}.

Following \citet{Ginzburg94} the state of the network of $N$ binary
model neurons is described by a binary vector $\n\in\{0,1\}^{N}$
and each neuron is updated at independently drawn time points with
exponentially distributed intervals of mean duration $\tau$. This
stochastic update constitutes a source of noise in the system. Given
the $i$-th neuron is updated, the probability to end in the up-state
($n_{i}=1$) is determined by the gain function $F_{i}(\n)$ which
depends on the activity $\n$ of all other neurons. The probability
to end in the down state ($n_{i}=0$) is $1-F_{i}(\n)$. Here we implemented
the binary model in the NEST simulator \citep{Gewaltig_07_11204}
as described in \nameref{sub:implementation_binary}. Such systems
have been considered earlier \citep{Ginzburg94,Buice09_377}, and
here we follow the notation employed in the latter work. In the following
we collect results that have been derived in these works and refer
the reader to these publications for the details of the derivations.
The zero-time lag covariance function is defined as $c_{ij}(t)=\langle n_{i}(t)n_{j}(t)\rangle-a_{i}(t)a_{j}(t)$,
with the expectation value $\langle\rangle$ taken over different
realizations of the stochastic dynamics. Here $\a(t)=(a_{1}(t),\ldots,a_{N}(t))^{T}$
is the vector of mean activities $a_{i}(t)=\langle n_{i}(t)\rangle$.
$c_{ij}(t)$ fulfills the differential equation

\begin{eqnarray*}
\tau\frac{d}{dt}c_{ij}(t) & = & -2c_{ij}(t)+\langle(n_{j}(t)-a_{j}(t))F_{i}(\n)\rangle+\langle(n_{i}(t)-a_{i}(t))F_{j}(\n)\rangle.
\end{eqnarray*}
In the stationary state, the correlation therefore fulfills
\begin{eqnarray}
c_{ij} & = & \frac{1}{2}\langle(n_{j}-a_{j})F_{i}(\n)\rangle+\frac{1}{2}\langle(n_{i}-a_{i})F_{j}(\n)\rangle.\label{eq:cij_stationary}
\end{eqnarray}
The time lagged covariance $c_{ij}(t,s)=\langle n_{i}(t)n_{j}(s)\rangle-a_{i}(t)a_{j}(s)$
fulfills for $t>s$ the differential equation
\begin{eqnarray}
\tau\frac{d}{dt}c_{ij}(t,s) & = & -c_{ij}(t,s)+\langle F_{i}(\n,t)(n_{j}(s)-a_{j}(s))\rangle.\label{eq:diffeq_cross_cov_lagged}
\end{eqnarray}
This equation is also true for $i=j$, the autocovariance. The term
$\langle F_{i}(\n,t)(n_{j}(s)-a_{j}(s))\rangle$ has a simple interpretation:.
iIt measures the influence of a fluctuation of neuron $j$ at time
$s$ around its mean value on the gain of neuron $i$ at time $t$
\citep{Ginzburg94}. We now assume a particular form for the coupling
between neurons
\begin{eqnarray}
F_{i}(\n,t) & = & \phi(\J{}_{i}\n(t-d))=\phi(\sum_{k=1}^{N}J{}_{ik}n_{k}(t-d)),\label{eq:input_functional}
\end{eqnarray}
where $\J_{i}$ is the vector of incoming synaptic weights into neuron
$i$ and $\phi$ is a non-linear gain function. Assuming that the
fluctuations of the total input $\J_{i}\n$ into the $i$-th neuron
are sufficiently small to allow a linearization of the gain function
$\phi$, we obtain the Taylor expansion
\begin{eqnarray*}
F_{i}(\n,t) & = & F_{i}(\a)+\phi^{\prime}(\J_{i}\a)\,\J_{i}(\n(t-d)-\a(t-d)),
\end{eqnarray*}
where 
\begin{align}
\phi^{\prime}(\J_{i}\a)\label{eq:slope_of_avg}
\end{align}
 is the slope of the gain function at the point of mean input.

\begin{figure}
\centering{}\includegraphics{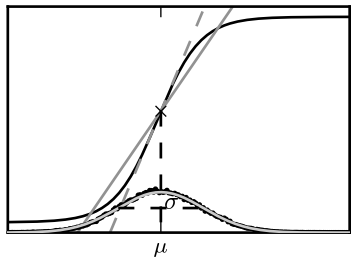}\caption{Alternative linearizations of the binary neuron model. The black curve
represents the non-linear gain function $\phi(x)=\frac{1}{2}+\frac{1}{2}\tanh(\beta x)$.
The dashed gray line is its tangent at the mean input value (denoted
by the diagonal cross). The solid curve is the slope $\langle\phi^{\prime}\rangle$
averaged over the distribution of the fluctuating input \eqref{eq:avg_slope}.
This distribution estimated from direct simulation is presented by
black dots, the corresponding theoretical prediction of a normal distribution
$\mathcal{N}(\mu,\sigma)$ \eqref{eq:avg_slope} is shown as the light
gray curve.\label{fig:linofbin}}
\end{figure}

Up to this point the treatment of the system is identical to the work
of \citet{Ginzburg94}. Now we present an alternative approach for
the linearization which takes into account the effect of fluctuations
in the input. For sufficiently asynchronous network states, the fluctuations
in the input $\J_{i}\n(t-d)$ to neuron $i$ can be approximated by
a Gaussian distribution $\mathcal{N}(\mu,\sigma)$. In the following
we consider a homogeneous random network with fixed in-degree as described
in \nameref{sub:population_averaged_correlations}. As each neuron
receives the same number $K$ of excitatory and $\gamma K$ inhibitory
synapses, the marginal statistics of the summed input to each neuron
is identical. The mean input to a neuron then is $\mu=KJ(1-\gamma g)a$,
where $a$ is the mean activity of a neuron in the network. If correlations
are small, the variance of this input signal distribution can be approximated
as the sum of the variances of the individual contributions from the
incoming signals, resulting in $\sigma^{2}=KJ^{2}(1+\gamma g^{2})\, a(1-a)$,
where we used the fact that the variance of a binary variable with
mean $a$ is $a(1-a)$. This results from a direct calculation: since
$n\in\{0,1\}$, $n^{2}=n$, so that the variance is $\langle n^{2}\rangle-\langle n\rangle^{2}=\langle n\rangle-\langle n\rangle^{2}=a(1-a)$.
Averaging the slope $\phi^{\prime}$ of the gain function over the
distribution of the input variable results in the averaged slope
\begin{align}
\langle\phi^{\prime}\rangle & \simeq\int_{-\infty}^{\infty}\mathcal{N}(\mu,\sigma,x)\,\phi^{\prime}(x)\; dx\label{eq:avg_slope}\\
\text{with }\mathcal{N}(\mu,\sigma,x) & =\frac{1}{\sqrt{2\pi}\sigma}\exp\left(-\frac{(x-\mu)^{2}}{2\sigma^{2}}\right).\nonumber 
\end{align}
The two alternative methods of linearization of $\phi$ are illustrated
in \prettyref{fig:linofbin}. In the given example, the linearization
procedure taking into account the fluctuations of the input signal
results in a smaller effective slope $\langle\phi^{\prime}\rangle$
than taking the slope $\phi^{\prime}(a)$ at the mean activity $a$
near its maximum. Averaging the slope $\langle\phi^{\prime}\rangle$
over this distribution fits simulation results better than $\phi^{\prime}(a)$
calculated at the mean of $a$, as shown in \prettyref{fig:binandlin}.

The finite slope of the non-linear gain function can be understood
as resulting from the combination of a hard threshold with an intrinsic
local source of noise. The inverse strength of this noise determines
the slope parameter $\beta$ \citep{Ginzburg94}. In this sense, the
network model contains two sources of noise, the explicit local noise,
quantified by $\beta$ and the fluctuating synaptic input interpreted
as self-generated noise on the network level, quantified by $\sigma$.
Even in the absence of local noise ($\beta\to\infty$), the above
mentioned linearization is applicable and yields a finite effective
slope $\langle\phi^{\prime}\rangle$ \eqref{eq:avg_slope}. In the
latter case the resulting effective synaptic weight is independent
of the original synapse strength \citep{Grytskyy13_258}.

We now extend the classical treatment of covariances in binary networks
\citep{Ginzburg94} by synaptic conduction delays. In \eqref{eq:input_functional}
$F_{i}(\n,t$) must therefore be understood as a functional acting
on the function $\n(t^{\prime})$ for $t^{\prime}\in[-\infty,t]$,
so that also synaptic connections with time delay $d$ can be realized.
We define an effective weight vector to absorb the gain factor as
$\w_{i}=\beta_{i}\J{}_{i}$, with either $\beta_{i}=\phi^{\prime}(\mu)$
or $\beta_{i}=\langle\phi^{\prime}\rangle$ depending on the linearization
procedure, and expand the right hand side of \prettyref{eq:diffeq_cross_cov_lagged}
to obtain
\begin{eqnarray*}
\langle F_{i}(\n,t)(n_{j}(s)-a_{j}(s))\rangle & = & \sum_{k=1}^{N}w_{ik}c_{kj}(t-d,s).
\end{eqnarray*}
Thus the cross-covariance fulfills the matrix delay differential equation
\begin{eqnarray}
\tau\frac{d}{dt}\c(t,s)+\c(t,s) & = & \w\c(t-d,s).\label{eq:diffeq_cross_cov}
\end{eqnarray}
This differential equation is valid for $t>s$. For the stationary
solution, the differential equation only depends on the relative timing
$u=t-s$ 
\begin{eqnarray}
\tau\frac{d}{du}\c(u)+\c(u) & = & \w\c(u-d).\label{eq:diffeq_cross_cov_eq}
\end{eqnarray}
The same linearization applied to \eqref{eq:cij_stationary} results
in the boundary condition for the solution of the previous equation

\begin{equation}
2\c(0)=\w\c(-d)+(\w\c(-d))^{T}
\end{equation}
 or, if we split $\c$ into its diagonal and its off-diagonal parts
$\c_{a}$ and $\c_{\neq}$

\begin{align}
2\c_{\neq}(0) & =\w\c_{\neq}(-d)+(\w\c_{\neq}(-d))^{T}+\OO\label{eq:binary_cov_0_lag}\\
\text{with } & \OO=\w\c_{a}(-d)+(\w\c_{a}(-d))^{T}.\nonumber 
\end{align}
In the following section we use this representation to demonstrate
the equivalence of the covariance structure of binary networks to
the solution for Ornstein-Uhlenbeck processes with input noise.

\subsection{Equivalence of binary neurons and Ornstein-Uhlenbeck processes\label{sub:equivalence_binary_oup}}

\begin{figure}
\begin{centering}
\includegraphics{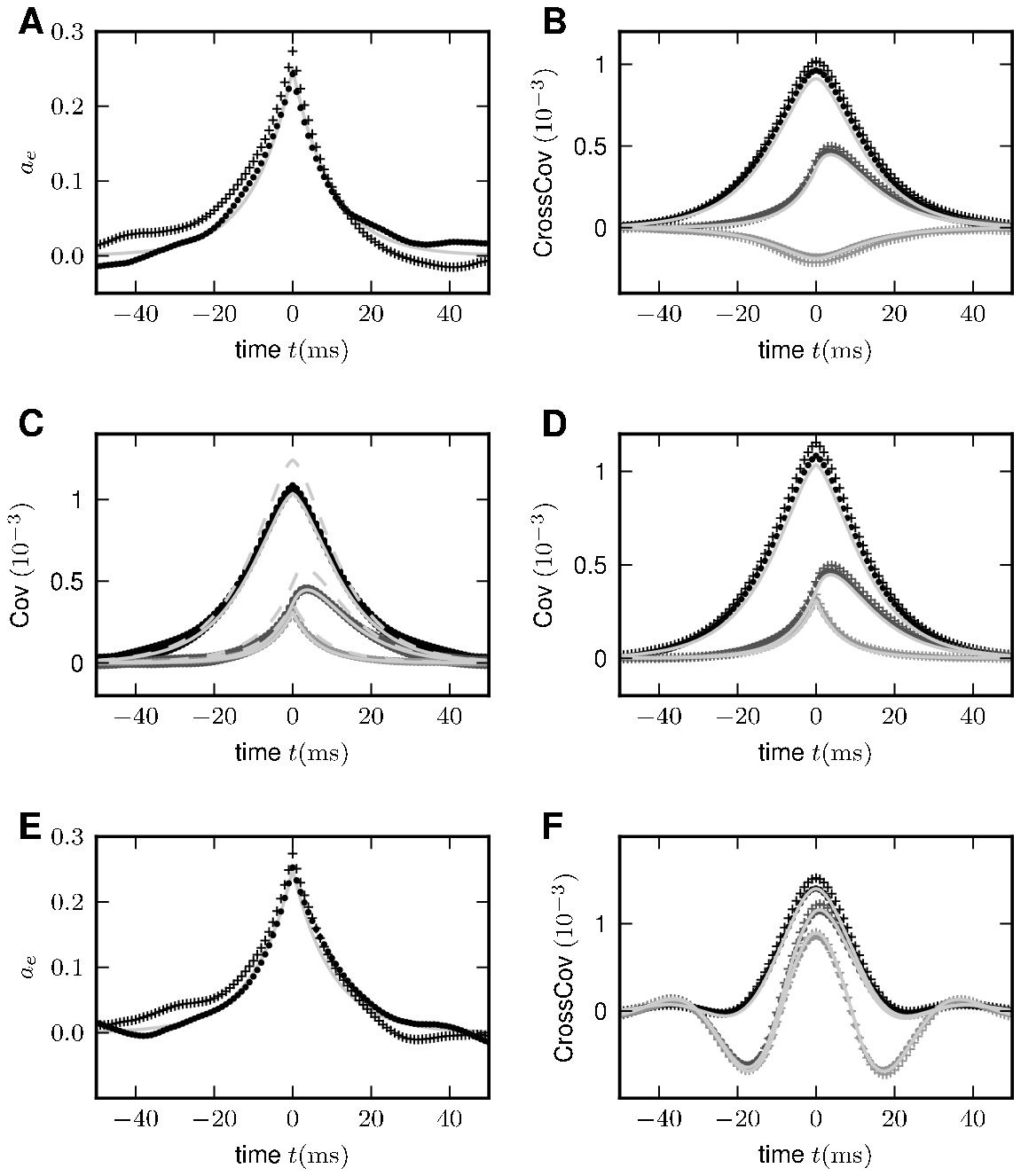}
\par\end{centering}

\caption{Binary model neuron corresponds to OUP model with input noise. Autocovariance
(\textbf{A}), crosscovarince (\textbf{B}), and autocovariance of population
averaged activity (\textbf{C},\textbf{D}) for binary neurons (dots)
and rate model with input noise (crosses). $c_{\Ex\Ex}$,$c_{\Ex\In}$
and $c_{\In\In}$ are shown in black, gray, and light gray. Corresponding
theoretical predictions (\eqref{eq:cov_avg_oupin_t} in C and D, \eqref{eq:auto_cov_in}
in A, their difference in C) are plotted as light gray curves throughout.
Dashed curve in C represents the theoretical prediction using the
linearization with the slope at the mean activity \eqref{eq:slope_of_avg},
the solid curve shows the results for the slope averaged over Gaussian
distributed input fluctuations \eqref{eq:avg_slope}. The spread of
the simulation results for binary neurons in panel C is due to different
realizations of the random connectivity. Panels (\textbf{E},\textbf{F})
are the same as (A,B) but for the presence of a synaptic delay $d=10\ms$
instead of $d=0.1\ms$.\label{fig:binandlin}}
\end{figure}

In the following subsection we show that the same equations \eqref{eq:diffeq_cross_cov_eq}
and \eqref{eq:binary_cov_0_lag} for binary neurons also hold for
the Ornstein-Uhlenbeck process (OUP) with input noise. In doing so
here we also extend the existing framework of Ornstein-Uhlenbeck processes
\citep{Risken96} to synaptic conduction delays $d$. A network of
such processes is described by

\begin{equation}
\tau\frac{d}{dt}\r(t)+\r(t)=\w\r(t-d)+\x(t),\label{eq:OUPin}
\end{equation}
where $\x$ is a vector of pairwise uncorrelated white noise with
$\langle\x(t)\rangle_{x}=0$ and $\langle x_{i}(t)x_{j}(t+t^{'})\rangle_{x}=\delta_{ij}\delta(t^{'})\rho^{2}$.
With the help of the Green's function $G$ satisfying $(\tau\frac{d}{dt}+1)\, G(t)=\delta(t)$,
namely $G(t)=\frac{1}{\tau}\,\theta(t)\, e^{-t/\tau}$, we obtain
the solution of equation \eqref{eq:OUPin} as

\[
\r(t)=\tau G(t)\r(0)+\int_{0}^{t}G(t-t^{'})(\w\r(t^{'}-d)+\x(t^{'}))\, dt^{'}.
\]
The equation for the fluctuations $\delta\r(t)=\r(t)-\langle\r(t)\rangle_{x}$
around the expectation value

\begin{align*}
\delta\r(t) & =\int_{0}^{t}G(t-t^{'})(\w\delta\r(t^{'}-d)+\x(t^{'}))\, dt^{'}
\end{align*}
coincides with the noisy rate model with input noise \eqref{eq:rdin}
with delay $d$ and convolution kernel $h=G$. In the next step we
investigate the covariance matrix $c_{ij}(t,s)=\langle\delta r_{i}(t+s)\delta r_{j}(t)\rangle_{x}$
to show for which choice of parameters the covariance matrices for
the binary model and the OUP with input noise coincide. To this end
we derive the differential equation with respect to the time lag $s$
for positive lags $s>0$

\begin{align}
\tau\frac{d}{ds}\c(t,s) & =\langle\tau\frac{d}{ds}\delta\r(t+s)\delta\r^{T}(t)\rangle{}_{x}\label{eq:masterforoupin}\\
 & =\langle(\w\delta\r(t+s-d)-\delta\r(t+s)+\x(t+s))\delta\r^{T}(t)\rangle{}_{x}\nonumber \\
 & =\w\c(t,s-d)-\c(t,s),\nonumber 
\end{align}
where we used $\langle\x(t+s))\delta\r(t)\rangle_{x}=0$, because
the noise is realized independently for each time step and the system
is causal. Eq. \eqref{eq:masterforoupin} is identical to the differential
equation satisfied by the covariance matrix \eqref{eq:diffeq_cross_cov}
for binary neurons \citep{Ginzburg94}. To determine the initial condition
of \eqref{eq:masterforoupin} we need to take the limit $\c(t,0)=\lim_{s\rightarrow+0}\c(t,s)$.
This initial condition can be obtained as the stationary solution
of the following differential equation

\begin{align*}
\tau\frac{d}{dt}\c(t,0) & =\lim_{s\rightarrow+0}(\langle\tau\frac{d}{dt}\delta\r(t+s)\delta\r^{T}(t)\rangle_{x}+\langle\delta\r(t+s)\tau\frac{d}{dt}\delta\r^{T}(t)\rangle_{x})\\
 & =\lim_{s\rightarrow+0}\left(\langle(\w\delta\r(t+s-d)-\delta\r(t+s)+\x(t+s))\delta\r^{T}(t)\rangle_{x}\right.\\
 & \phantom{=\lim_{s\rightarrow+0}(}\left.+\langle\delta\r(t+s)(\delta\r^{T}(t-d)\w^{T}-\delta\r^{T}(t)+\x^{T}(t))\rangle_{x}\right)\\
 & =-2\c(t,0)+\w\c(t,-d)+\c(t-d,d)\w^{T}+\D.
\end{align*}
Here we used that $\langle\x(t+s)\delta\r^{T}(t)\rangle$ vanishes
due to independent noise realizations and causality and

\begin{align*}
\D & =\lim_{s\rightarrow+0}\langle\delta\r(t+s)\x^{T}(t)\rangle_{x}\\
 & =\lim_{s\rightarrow+0,\ s<d}\int_{0}^{t+s}G(t+s-t^{'})(\w\underbrace{\langle\delta\r(t^{'}-d)\x^{T}(t)\rangle_{x}}_{=0\text{ causality}}+\underbrace{\langle\x(t^{'})\x^{T}(t)\rangle_{x}}_{=\Em\delta(t-t^{\prime})\rho^{2}})dt^{'}\\
 & =\lim_{s\rightarrow+0,\ s<d}\int_{0}^{t+s}G(t+s-t^{'})\Em\delta(t-t^{'})\rho^{2}dt^{'}\\
 & =\lim_{s\rightarrow+0,\ s<d}G(s)\Em\sigma^{2}=\frac{1}{\tau}\Em\rho^{2}.
\end{align*}
In the stationary state, $\c$ only depends on the time lag $s$ and
is independent of the first time argument $t$, which, with the symmetry
$\c(-d)^{T}=\c(d)$ yields the additional condition for the solution
of \eqref{eq:masterforoupin}

\[
2\c(0)=\w\c(-d)+(\w\c(-d))^{T}+\D
\]
or, if $\c$ is split in diagonal and off-diagonal parts $\c_{a}$
and $\c_{\neq}$, respectively,

\[
\begin{array}{cc}
2\c_{\neq}(0) & =\w\c_{\neq}(-d)+(\w\c_{\neq}(-d))^{T}+\OO\\
2\c_{a}(0) & =\w\c_{\neq}(-d)+(\w\c_{\neq}(-d))^{T}+\D
\end{array}
\]
with \foreignlanguage{english}{\textrm{$\OO=\w\c_{a}(-d)+(\w\c_{a}(-d))^{T}$}}.
In the equation for the autocovariance $\c_{a}$ the first two terms
are contributions due to the cross covariance. In the state of asynchronous
network activity with $c_{ij}\sim N^{-1}\ \text{for }\ i\neq j$ these
terms are typically negligible in comparison to the third term because
$\sum_{k}w_{ik}c_{ki}\sim wKN^{-1}=pw$, which is typically smaller
than $1$ for small effective weights $w<1$ and small connection
probabilities $p\ll1$. In this approximation with \eqref{eq:masterforoupin}
the temporal shape of the autocovariance function is exponentially
decaying with time constant $\tau$. With $\c_{a}(0)\approx\D/2$
the approximate solution for the autocovariance is
\begin{align}
\c_{a}(t) & =\frac{\D}{2}\exp(-\frac{|t|}{\tau}).\label{eq:auto_cov_in}
\end{align}
The cross covariance then satisfies the initial condition

\begin{align*}
2\c_{\neq}(0) & =\w\c_{\neq}(-d)+(\w\c_{\neq}(-d))^{T}+\OO\\
\OO & =\w\D/2+(\w\D/2)^{T},
\end{align*}
which coincides with \eqref{eq:binary_cov_0_lag} for binary neurons
if the diagonal matrix containing the zero time autocorrelations $\c_{a}(0)$
for binary neurons is equal to $\D/2$, i.e. if the amplitude of the
input noise $\rho^{2}=2\tau a(1-a)$ and the effective linear coupling
satisfies $\w_{i}=\beta_{i}\J{}_{i}$. \prettyref{fig:binandlin}
shows simulation results for population averaged covariance functions
in binary networks and in networks of OUPs with input noise where
the parameters of the OUP network are chosen according to the requirements
derived above. The theoretical results \eqref{eq:cov_avg_oupin_t}
 agree well with the direct simulations of both systems. For comparison,
both methods of linearization, as explained above, are shown. The
linearization procedure which takes into account the noise on the
input side of the non-linear gain function results in a more accurate
prediction. Moreover, the results derived here extend the classical
theory \citep{Ginzburg94} by considering synaptic conduction delays.
\prettyref{fig:echoes} shows the decomposition of the covariance
structure for a non-zero delay $d=3\ms$. For details of the implementation
see \nameref{sub:implementation_binary}. The explicit effect of
introducing delays into the system, such as the appearance of oscillations
in the time dependent covariance, is presented in panels E and F of
\prettyref{fig:binandlin}, differing from panels A and B of this
figure, respectively, only in the delay ($d=10\ms$ for E and F, $d=0.1\ms$
for A and B).

\section{Hawkes processes\label{sec:equivalence_hawkes_oup}}

In the following section we show that to linear order the covariance
functions in networks of Hawkes processes \citep{Hawkes71_438} are
equivalent to those in the linear rate network with output noise.
Hawkes processes generate spikes randomly with a time density given
by $\r(t)$, where neuron $i$ generates spikes at a rate $r_{i}(t)$,
realized independently within each infinitesimal time step. Arriving
spike trains $\s$ influence $\r$ according to 
\begin{align}
\r(t) & =\nu+(h_{d}\ast\J\s)(t),\label{eq:hawkes_def}
\end{align}
with the connectivity matrix $\J$ and the kernel function $h_{d}$
including the delay. Here $\nu$ is a constant base rate of spike
emission assumed to be equal for each neuron. Here we employ the implementation
of the Hawkes model in the NEST simulator \citep{Gewaltig_07_11204}.
The implementation is described in \nameref{sub:implementation_hawkes}.

Given neuron $j$ spiked at time $u\leq t$, the probability of a
spike in the interval $[t,t+\delta t)$ for neuron $i$ is $1$ if
$i=j,\ u=t$ (the neuron spikes synchronously with itself) and $r_{i}(t)\delta t+o(\delta t^{2})$
otherwise. Considering the system in the stationary state with the
time averaged activity $\bar{\r}=\langle\s(t)\rangle$ we obtain a
convolution equation for time lags $\tau\ge0$ for the covariance
matrix with the entry $c_{ij}(\tau)$ for the covariance between spike
trains of neurons $i$ and $j$

\begin{align}
\c(\tau) & =\langle\s(t+\tau)\s^{T}(t)\rangle-\langle\s(t+\tau)\rangle\langle\s^{T}(t)\rangle\label{eq:convolution_eq_hawkes}\\
 & =\langle(\delta(\tau)\Em+\r(t+\tau))\s^{T}(t)\rangle-\bar{\r}\bar{\r}^{T}\nonumber \\
 & =\langle\r(t+\tau)(\s^{T}(t)-\bar{\r}^{T})\rangle+\D_{\overline{\r}}\nonumber \\
 & =\langle(\nu+(h_{d}\ast\J\s)(t+\tau))(\s^{T}(t)-\bar{\r}^{T})\rangle+\D_{\overline{\r}}\nonumber \\
 & =h_{d}\ast\J\langle\s(t+\tau)(\s^{T}(t)-\bar{\r}^{T})\rangle+\D_{\overline{\r}}\nonumber \\
 & =(h_{d}\ast\J\c)(\tau)+\D_{\overline{\r}},\nonumber 
\end{align}
with the diagonal matrix $\D_{\overline{\r}}=\delta(\tau)\diag(\bar{\r})$,
which has been derived earlier \citep{Hawkes71_438}. If the rates
of all neurons are equal, $\bar{\r}_{i}=\bar{r}$, all entries in
the diagonal matrix are the same, $\D_{\overline{\r}}=\delta(\tau)\mathbf{\Em}\bar{r}$.
In the subsequent section we demonstrate that the same convolution
equation \eqref{eq:convolution_eq_hawkes} holds for the linear rate
with output noise.

\subsection{Convolution equation for linear noisy rate neurons\label{sub:convolution_equation_OUP}}

For the linear rate model with output noise we use equation \eqref{eq:rdon}
for time lags $\tau>0$ to obtain a convolution equation for the covariance
matrix of the output signal vector $\y=\r+\x$ as

\begin{align}
\c(\tau) & =\langle\y(t+\tau)(\y^{T}(t)-\bar{\r}^{T})\rangle\label{eq:convolution_eq_oup_on}\\
 & =\langle(h_{d}\ast\w\y+\x)(t+\tau)(\y^{T}(t)-\bar{\r}^{T})\rangle\nonumber \\
 & =(h_{d}\ast\w\c)(\tau)+\langle\x(t+\tau)(\r^{T}(t)-\bar{\r}^{T})\rangle+\langle\x(t+\tau)\x^{T}(t)\rangle\nonumber \\
 & =(h_{d}\ast\w\c)(\tau)+\D,\nonumber 
\end{align}
where we utilized that due to causality the random noise signal generated
at $t+\tau$ has no influence on $\r(t)$, so the respective correlation
vanishes. $\D$ is the covariance of the noise as in \eqref{eq:cross_spectrum_on},
$D_{ij}(\tau)=\langle x_{i}(t)x{}_{j}(t+\tau)\rangle=\delta_{ij}\delta(\tau)\rho^{2}$.
If $\rho$ is chosen such that $\rho^{2}$ coincides with the averaged
activity $\bar{r}$ in a network of Hawkes neurons and the connection
matrix $\w$ is identical to $\J$ of the Hawkes network, the equations
\eqref{eq:convolution_eq_hawkes} and \eqref{eq:convolution_eq_oup_on}
are identical. Therefore the cross spectrum of both systems is given
by \eqref{eq:cross_spectrum_on}.

\subsection{Non-linear self-consistent rate in rectifying Hawkes networks}

The convolution equation \eqref{eq:convolution_eq_hawkes} for the
covariance matrix of Hawkes neurons is exact if no element of $\r$
is negative, which is particularly the case for a network of only
excitatory neurons. Especially in networks including inhibitory couplings,
the intensity $r_{i}$ of neuron $i$ may assume negative values.
A neuron with $r_{i}<0$ does not emit spikes, so the instantaneous
rate is given by $\lambda_{i}=[r_{i}(t)]_{+}=\theta(r_{i}(t))\, r_{i}(t),$
with the Heaviside function $\theta$. We now take into account this
effective nonlinearity --the rectification of the Hawkes model neuron--
in a similar manner as we already used to linearize binary neurons.
If the network is in the regime of low spike rates, the fluctuations
in the input of each neuron due to the Poissonian arrival of spikes
are large compared to the fluctuations due to the time varying intensities
$\r(t)$. Considering the same homogeneous network structure as described
in \nameref{sub:population_averaged_correlations}, the input statistics
is identical for each cell $i$, so the mean activity $\lambda_{0}=\left\langle \lambda_{i}\right\rangle $
is the same for all neurons $i$. The superposition of the synaptic
inputs to neuron $i$ cause an instantaneous intensity $r_{i}$ that
follows approximately a Gaussian distribution $\mathcal{N}(\mu,\sigma,r_{i})$
with mean $\mu=\langle r\rangle=\nu+\lambda_{0}KJ(1-g\gamma)$ and
standard deviation $\sigma=\sqrt{\langle r^{2}\rangle-\langle r\rangle^{2}}=J\sqrt{\frac{\lambda_{0}}{2\tau}K(1+g^{2}\gamma)}$.
These expressions hold for the exponential kernel \eqref{eq:exp_kernel-1}
due to Campbell's theorem \citep{PapoulisProb4th}, because of the
stochastic Poisson-like arrival of incoming spikes, where the standard
deviation of the spike count is proportional to the square root of
the intensity $\lambda_{0}$. The rate $\lambda_{0}$ is accessible
by explicit integration over the Gaussian probability density as
\begin{align*}
\lambda_{0} & =\int_{-\infty}^{\infty}\mathcal{N}(\mu,\sigma,r)\, r\,\theta(r)\, dr\\
 & =\frac{1}{\sqrt{2\pi}\sigma}\int_{0}^{\infty}\exp(-\frac{(r-\mu)^{2}}{2\sigma^{2}})\, r\, dr\\
 & =\frac{-\sigma}{\sqrt{2\pi}}\int_{0}^{\infty}\exp(-\frac{(r-\mu)^{2}}{2\sigma^{2}})\,\frac{-(r-\mu)}{\sigma^{2}}\, dr+\frac{\mu}{\sqrt{2\pi}\sigma}\int_{0}^{\infty}\exp(-\frac{(r-\mu)^{2}}{2\sigma^{2}})\, dr\\
 & =\frac{\sigma}{\sqrt{2\pi}}\exp(-\frac{\mu^{2}}{2\sigma^{2}})+\frac{\mu}{2}(1-\text{erf}(-\frac{\mu}{\sqrt{2}\sigma})).
\end{align*}
This equation needs to be solved self-consistently (numerically or
graphically) to determine the rate in the network, as the right hand
side depends on the rate $\lambda_{0}$ itself through $\mu$ and
$\sigma$. Rewritten as

\begin{align}
\lambda_{0} & =\frac{\sigma}{\sqrt{2\pi}}\exp(-\frac{\mu^{2}}{2\sigma^{2}})+\mu P_{\mu,\sigma}(r>0)\nonumber \\
P_{\mu,\sigma}(r>0) & =\frac{1}{2}-\frac{1}{2}\text{erf}(-\frac{\mu}{\sqrt{2}\sigma}),\label{eq:prob_positive_rate_Hawkes}
\end{align}
$P_{\mu,\sigma}(r>0)$ is the probability that the intensity of a
neuron is above threshold and therefore contributes to the transmission
of a small fluctuation in the input. A neuron for which $r<0$ acts
as if it was absent. Hence we can treat the network with rectifying
neurons completely analogous to the case of linear Hawkes processes,
but multiply the synaptic weight $J$ or $-gJ$ of each neuron with
$P_{\mu,\sigma}(r>0)$, i.e. the linearized connectivity matrix is 

\begin{equation}
\w=P_{\mu,\sigma}(r>0)\J.\label{eq:hawkes_lin_weight}
\end{equation}
  \prettyref{fig:covariance_spiking} shows the agreement of the
covariance functions obtained from direct simulation of the network
of Hawkes processes and the analytical solution \eqref{eq:cov_avg_oupon_t}
with average firing rate $\lambda_{0}$ determined by \eqref{eq:prob_positive_rate_Hawkes},
setting the effective strength of the noise $\rho^{2}=\lambda_{0}$,
and the linearized coupling as described above. The detailed procedure
for choosing the parameters in the direct simulation is described
together with the implementation of the Hawkes model in \nameref{sub:implementation_hawkes}.

\section{Leaky integrate-and-fire neurons\label{sec:equivalence_lif_oup}}

\begin{figure}
\begin{centering}
\includegraphics{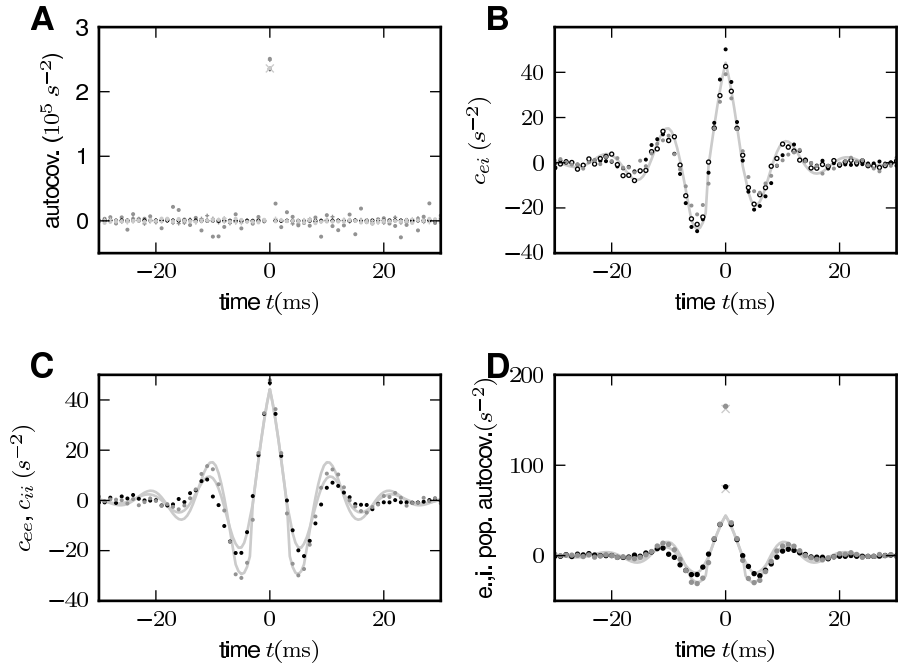}
\par\end{centering}

\caption{Covariance structure in spiking networks corresponds to OUP with output
noise. \textbf{A} Autocovariance obtained by direct simulation of
the LIF (black), Hawkes (gray), and OUP (light gray) models for excitatory
(dots) and inhibitory neurons (crosses). \textbf{B} Covariance $c_{\Ex\In}$
averaged over disjoint pairs of neurons for LIF (black dots), Hawkes
(gray dots), and OUP with output noise (empty circles). \textbf{C}
Covariance averaged over disjoint pairs of neurons of the same type.\textbf{
D} Autocovariance of the population averaged activity. Averages in
C,D over excitatory neurons as black dots, over inhibitory neurons
as gray dots. Corresponding theoretical predictions \eqref{eq:cov_avg_oupon_t}
are plotted as light gray curves in all panels except A. Light gray
diagonal crosses in A and D denote theoretical peak positions determined
by the firing rate $\bar{r}$ as $\bar{r}\Delta t$ (where $\Delta t=0.1\ms$
is the time resolution of the histogram).\label{fig:covariance_spiking} }
\end{figure}

In this section we consider a network of leaky integrate-and-fire
(LIF) model neurons with exponentially decaying postsynaptic currents
and show its equivalence to the network of Ornstein-Uhlenbeck processes
with output noise, valid in the asynchronous irregular regime. A spike
sent by neuron $j$ at time $t$ arrives at the target neuron $i$
after the synaptic delay $d$, elicits a synaptic current $I_{i}$
that decays with time constant $\taus$ and causes a response in the
membrane potential $V_{i}$ proportional to the synaptic efficacy
$J_{ij}$. With the time constant $\taum$ of the membrane potential,
the coupled set of differential equations governing the subthreshold
dynamics of a single neuron $i$ is \citep{Fourcaud02}
\begin{eqnarray}
\taum\frac{dV_{i}}{dt} & = & -V_{i}+I_{i}(t)\nonumber \\
\taus\frac{dI_{i}}{dt} & = & -I_{i}+\taum\sum_{j=1,j}^{N}J_{ij}s_{j}(t-d),\label{eq:diffeq_iaf}
\end{eqnarray}
where the membrane resistance was absorbed into the definitions of
$J_{ij}$ and $I_{i}$. If $V_{i}$ reaches the threshold $V_{\theta}$
at time point $t_{k}^{i}$ the neuron emits an action potential and
the membrane potential is reset to $V_{r}$, where it is clamped for
the refractory time $\taur$. The spiking activity of neuron $i$
is described by this sequence of action potentials, the spike train
$s_{i}(t)=\sum_{k}\delta(t-t_{k}^{i})$. The dynamics of a single
neuron is deterministic, but in network states of asynchronous, irregular
activity and in the presence of external Poisson inputs to the network,
the summed input to each cell can well be approximated as white noise
\citep{Brunel00_183} with first moment $\mu_{i}=\taum\sum_{j}J_{ij}r_{j}$
and second moment $\sigma_{i}^{2}=\taum\sum_{j}J_{ij}^{2}r_{j}$,
where $r_{j}$ is the stationary firing rate of neuron $j$. The stationary
firing rate of neuron $i$ is then given by \citep{Fourcaud02}
\begin{eqnarray}
r_{i}^{-1} & = & \taur+\taum\sqrt{\pi}\left(F(y_{\theta})-F(y_{r})\right)\label{eq:rate}\\
f(y) & = & e^{y^{2}}(1+\mathrm{erf}(y))\quad F(y)=\int^{y}f(y)\, dy\nonumber \\
\text{with }y_{\theta,r} & = & \frac{V_{\theta,r}-\mu_{i}}{\sigma_{i}}+\frac{\alpha}{2}\sqrt{\frac{\taus}{\taum}}\quad\alpha=\sqrt{2}|\zeta(\frac{1}{2})|,\nonumber 
\end{eqnarray}
with Riemann's zeta function $\zeta$. The response of the LIF neuron
to the injection of an additional spike into afferent $j$ determines
the impulse response $w_{ij}h(t)$ of the system. The time integral
$w_{ij}=w_{ij}\int_{0}^{\infty}h(t)\, dt$ is the DC-susceptibility,
which can formally be written as the derivative of the stationary
firing rate by the rate of the afferent $r_{j}$, which, evaluated
by help of \eqref{eq:rate}, yields \citep[ Results and App. A]{Helias13_023002}

\begin{eqnarray}
w_{ij} & = & \frac{\partial r_{i}}{\partial r_{j}}=\alpha J_{ij}+\beta J_{ij}^{2}\label{eq:w_ij}\\
\text{with }\alpha & = & \sqrt{\pi}(\taum r_{i})^{2}\frac{1}{\sigma_{i}}\left(f(y_{\theta})-f(y_{r})\right)\nonumber \\
\text{and }\text{\ensuremath{\beta}} & = & \sqrt{\pi}(\taum r_{i})^{2}\frac{1}{2\sigma_{i}^{2}}\left(f(y_{\theta})\,\frac{V_{\theta}-\mu_{i}}{\sigma_{i}}-f(y_{r})\,\frac{V_{r}-\mu_{i}}{\sigma_{i}}\right).\nonumber 
\end{eqnarray}
In the strongly fluctuation-driven regime, the temporal behavior of
the kernel $h$ is dominated by a single exponential decay, whose
time constant can be determined empirically. In a homogeneous random
network the firing rates of all neurons are identical $r_{i}=\bar{r}$
and follow from the numerical solution of the self-consistency equation
\eqref{eq:rate}. Approximating the autocovariance function of a single
spike train by a $\delta$-peak scaled by the rate $\bar{r}\delta(t)$,
one obtains for the covariance function $\c$ between pairs of spike
trains the same convolution equation \eqref{eq:convolution_eq_hawkes}
as for Hawkes neurons \citep[cf. eq. 5]{Helias13_023002}. As shown
in \nameref{sub:convolution_equation_OUP} this convolution equation
coincides with that of a linear rate model with output noise \eqref{eq:convolution_eq_oup_on},
where the diagonal elements of $\D$ are chosen to agree to the average
spike rate $\rho^{2}=\bar{r}$. The good agreement of the analytical
cross covariance functions \eqref{eq:cov_avg_oupon_t} for the OUP
with output noise and direct simulation results for LIF are shown
in \prettyref{fig:covariance_spiking}.

\section{Discussion}

In this work we describe the path to a unified theoretical view on
pairwise correlations in recurrent networks. We consider binary neuron
models, leaky integrate-and-fire models, and linear point process
models. These models containing a non-linearity (spiking threshold
in spiking models, non-linear sigmoidal gain function in binary neurons,
strictly positive rates in Hawkes processes) are linearized, taking
into account the distribution of the fluctuating input.

The work presents results for several neuron models: We derive analytical
expressions for delay-coupled Ornstein-Uhlenbeck processes with input
and with output noise, we extend the analytical treatment for stochastic
binary neurons to the presence of synaptic delays, present a method
that takes into account network-generated noise to determine the effective
gain function, extend the theory of Hawkes processes to the existence
of delays and inhibition, and present in eq. \eqref{eq:poles_z} a
condition for the onset of global oscillations caused by delayed feedback,
generalized to feedback pathways through different eigenvalues of
the connectivity.

Some results qualitatively extend the existing theory (delays, inhibition),
others improve the accuracy of existing theories (linearization including
fluctuations). More importantly, our approach enables us to demonstrate
the equivalence of each of these models after linear approximation
to a linear model with fluctuating continuous variables. The fact
that linear perturbation theory leads to effective linear equations
is of course not surprising, but the analytical procedure firstly
enables a mapping between models that conserves quantitative results
and secondly allows us to uncover common structures underlying the
emergence of correlated activity in recurrent networks. For the
commonly appearing exponentially decaying response kernel function,
these rate models coincide with the Ornstein-Uhlenbeck process \citep[OUP, ][]{Uhlenbeck30,Risken96}.
We find that the considered models form two groups, which, in linear
approximation merely differ by a matrix valued factor scaling the
noise and in the choice of variables interpreted as neural activity.
The difference between these two groups corresponds to the location
of the noise: spiking models -- leaky integrate-and-fire models and
Hawkes models -- belong to the class with noise on the output side,
added to the activity of each neuron. The non-spiking binary neuron
model corresponds to an OUP where the noise is added on the input
side of each neuron. The closed solution for the correlation structure
of OUP holds for both classes.

\begin{figure}
\begin{centering}
\includegraphics{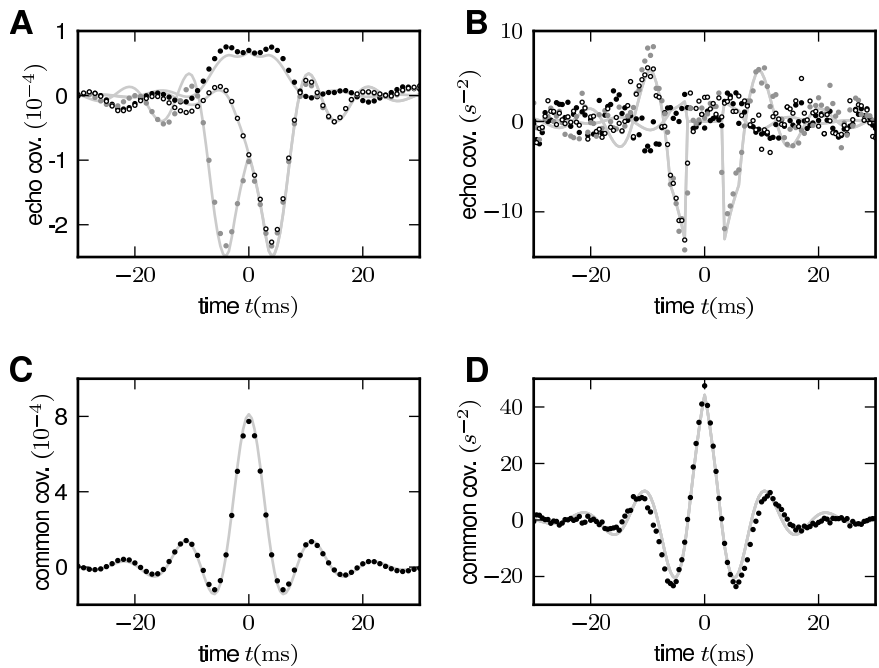}
\par\end{centering}

\caption{Different echo terms for spiking and non-spiking neurons. Binary non-spiking
neurons shown in A,C and LIF in B,D. \textbf{A},\textbf{B} Echo terms
by direct influence of the neuron's output on the network in dependence
of neuron types (in A, B $c_{\Ex\Ex}$,$c_{\Ex\In}$, and $c_{\In\In}$
are plotted as black, gray dots and circles). \textbf{C},\textbf{D}
Contributions to the covariance evoked by correlated and common input
(black dots) measured with help of auxiliary model neurons which do
not provide feedback to the network. Corresponding theoretical predictions
\eqref{eq:cross_spectrum_on_separated} are plotted as light gray
curves throughout.\label{fig:echoes}}
\end{figure}
We identify different contributions to correlations in recurrent networks:
the solution for output noise is split into three terms corresponding
to the $\delta$-peak in the autocovariance, the covariance caused
by shared input, and the direct synaptic influence of stochastic fluctuations
of one neuron on another --the latter echo terms are equal to propagators
acting with delays \citep{Helias13_023002}. A similar splitting into
echo and correlated input terms for the case of input noise is shown
in \prettyref{fig:echoes}. For increasing network size $N\rightarrow\infty$,
keeping the connection probability $p$ fixed, so that $K=pN$, and
with rescaled synaptic amplitudes $J\sim1/\sqrt{N}$ \citep{Vreeswijk96,Renart10_587}
the echo terms vanish fastest. Formally this can be seen from \eqref{eq:cov_avg_oupin_t}:
the multiplicative factor of the common covariance term $\varphi_{4}$
does not change with $N$ while the other coefficients decrease. So
ultimately all four entries of the matrix $\c$ have the same time
dependence determined by the common covariance term $\varphi_{4}$.
In particular the covariance between excitation and inhibition $c_{\Ex\In}$
becomes symmetric in this limit. This finally provides a quantitative
explanation of the observation made in \citep{Renart10_587} that
the time-lag between excitation and inhibition vanishes in the limit
of infinitely large networks. For a different synaptic rescaling $J\sim N^{-1}$
while keeping $\rho^{2}$ constant by appropriate additional input
to each neuron \citep[see ][ applied to the LIF model]{Helias13_023002},
all multiplicative factors decrease $\sim N^{-1}$ and so does the
amplitude of all covariances. Hence the asymmetry of $c_{\Ex\In}$
does not vanish in this limit. The same results hold for the case
of output noise where the term with $\varphi_{1}$ describes the common
input part of the covariance. In this case and for finite network
size, $c_{\In\Ex}$ coincides with $c_{\Ex\Ex}$ and $c_{\Ex\In}$
with $c_{\In\In}$ for $t>0$, having a discontinuous jump at the
time of the synaptic delay $t=d$. For time lags smaller than the
delay all four covariances coincide. This is due to causality, as
the second neuron cannot feel the influence of a fluctuation that
happened in the first neuron less than one synaptic delay before.
The covariance functions for systems corresponding to an OUP with
input noise contain neither discontinuities nor sharp peaks at $t=d$,
but $c_{\Ex\In}$ and $c_{\In\Ex}$ have maxima and minima near this
location. This observation can be interpreted as a result of the stochastic
nature of the binary model where changes in the input influence the
state of the neuron only with a certain probability. So, the entries
of $\c$ in this case take different values for $|t|<d$ but show
the tendency to approach each other with increasing $|t|\gg d$. This
tendency increases with network size. Our analytical solutions \eqref{eq:cov_avg_oupin_t}
for input noise and \eqref{eq:cov_avg_oupon_t} for output noise hence
explain the model-class dependent differences in the shape of covariance
functions.

The two above mentioned synaptic scaling procedures are commonly termed
{}``strong coupling'' ($J\sim1/\sqrt{N}$) and {}``weak coupling''
($J\sim1/N$), respectively. The results shown in \prettyref{fig:binandlin}
were obtained for $J=2/\sqrt{N}$ and $\beta=0.5$, so the number
of synapses required to cause a notable effect on the gain function
is $1/(\beta J)=\sqrt{N}$, which is small compared to the number
of incoming synapses $pN$. Hence the network is in the strong coupling
regime. Also note that for infinite slope of the gain function, $\beta\to\infty$,
the magnitude of the covariance becomes independent of the synaptic
amplitude $J$, in agreement with the linear theory presented here.
This finding can readily be understood by the linearization procedure,
presented in the current work, that takes into account the network-
generated fluctuations of the total input. The amplitude $\sigma$
of these fluctuations scales linearly in $J$ and the effective susceptibility
depends on $J/\sigma$ in the case $\beta\to\infty$, explaining the
invariance \citep{Grytskyy13_258}. In the current manuscript we generalized
this procedure to finite slopes $\beta$ and to other models than
the binary neuron model.

Our approach enables us to map results obtained for one neuron model
to another, in particular we extend the theory of all considered models
to capture synaptic conduction delays, and devise a simpler way to
obtain solutions for systems considered earlier \citep{Ginzburg94}.
Our derivation of covariances in spiking networks does not rely on
the advanced Wiener-Hopf method \citep{Hazewinkel02}, as earlier
derivations \citep{Hawkes71_438,Helias13_023002} do, but only employs
elementary methods. Our results are applicable for general connectivity
matrices, and for the purpose of comparison with simulations we explicitly
derive population averaged results. The averages of the dynamics of
the linear rate model equations are exact for random network architectures
with fixed out-degree, and approximate for fixed in-degree. Still,
for non-linear models the linearization for fixed in-degree networks
are simpler, because the homogeneous input statistics results in an
identical linear response kernel for all cells. Finally we show that
the oscillatory properties of networks of integrate-and-fire models
\citep{Brunel00_183,Helias13_023002} are model-invariant features
of all of the studied dynamics, given inhibition acts with a synaptic
delay. We relate the collective oscillations to the pole structure
of the cross spectrum, which also determines the power spectra of
population signals such as EEG, ECoG, and the LFP.

The presented results provide a further step to understand the shape
and to unify the description of correlations in recurrent networks.
We hope that our analytical results will be useful to constrain the
inverse problem of determining the synaptic connectivity given the
correlation structure of neurophysiological activity measurements.
Moreover the explicit expressions for covariance functions in the
time domain are a necessary prerequisite to understand the evolution
of synaptic amplitudes in systems with spike-timing dependent plasticity
and extend the existing methods \citep{Burkitt07_533,Gilson09_1,Gilson10_si}
to networks including inhibitory neurons and synaptic conduction delays.

\section{Conflict of Interest Statement}

The authors declare that the research was conducted in the absence
of any commercial or financial relationships that could be construed
as a potential conflict of interest.

\section{Appendix}

\subsection{Calculation of the population averaged cross covariance in time domain\label{sub:explicit_expressions_app}}

We obtain the population averaged cross spectrum for the Ornstein-Uhlenbeck
process with input noise by inserting the averaged connectivity matrix
$\w=\M$ \eqref{eq:averaged_conn} into \eqref{eq:cross_spectrum_in}.
The two eigenvalues of $\M$ are $0$ and $L=Kw(1-\gamma g)$. Taking
these into account, we first rewrite the term 

\begin{align*}
 & (H_{d}(\omega)^{-1}-\M)^{-1}\\
= & \det(H_{d}(\omega)^{-1}-\M)^{-1}\left(\begin{array}{cc}
H_{d}(\omega)^{-1}+Kw\gamma g & -Kw\gamma g\\
Kw & H_{d}(\omega)^{-1}-Kw
\end{array}\right)\\
= & ((H_{d}(\omega)^{-1}-0)(H_{d}(\omega)^{-1}-L))^{-1}\left(H_{d}(\omega)^{-1}\Em+Kw\left(\begin{array}{cc}
\gamma g & -\gamma g\\
1 & -1
\end{array}\right)\right)\\
= & f(\omega)\left(\Em+Kw\left(\begin{array}{cc}
\gamma g & -\gamma g\\
1 & -1
\end{array}\right)H_{d}(\omega)\right),
\end{align*}
where we introduced $f(\omega)=(H_{d}(\omega)^{-1}-L){}^{-1}$. The
corresponding transposed and conjugate complex term follows analogously.
Hence we obtain the expression for the cross spectrum \eqref{eq:avg_cross_spectrum_in}.
The residue of \foreignlanguage{english}{\textrm{$f(\omega)$}} at
$\omega=z_{k}(L)$ is 
\begin{align*}
\mathrm{Res}(f,\omega=z_{k}(L)) & =\lim_{\omega_{1}\rightarrow\omega}\frac{\omega_{1}-\omega}{f^{-1}(\omega_{1})}\\
 & \stackrel{\text{l'Hopital}}{=}\lim_{\omega_{1}\rightarrow\omega}\frac{1}{(f^{-1})^{\prime}(\omega_{1})}=\left(\frac{d(e^{i\omega d}(1+i\omega\tau))}{d\omega}\right)^{-1}\\
 & =\left(ide^{i\omega d}(1+i\omega\tau)+i\tau e^{i\omega d}\right)^{-1}=\left(idL+i\tau e^{i\omega d}\right)^{-1},
\end{align*}
where in the last step we used the condition for a pole $H_{d}(z_{k})^{-1}=e^{iz_{k}d}(1+iz_{k}\tau)=L$
(see \nameref{sub:spectrum_of_dynamics}). The residue of $H_{d}(\omega)$
at $z(0)=\frac{i}{\tau}$ is $-\frac{i}{\tau}e^{d/\tau}$. Using the
residue theorem, we need to sum over all poles within the integration
contour $\left\{ z_{k}(L)|k\in\mathbb{N}\right\} \cup\frac{i}{\tau}$
to get the expression for $\c(t)=\frac{1}{2\pi}\int_{-\infty}^{+\infty}\C(\omega)e^{i\omega t}d\omega=i\sum_{z\in\left\{ z_{k}(L)|k\in\mathbb{N}\right\} \cup\frac{i}{\tau}}\mathrm{Res}(\C(z),z)e^{izt}$
for $t\geqq0$. Sorting \eqref{eq:avg_cross_spectrum_in} to obtain
four matrix prefactors and remainders with different frequency dependence,
$\Phi_{1}(\omega)=f(\omega)f(-\omega)$, $\Phi_{2}(\omega)=f(\omega)f(-\omega)H_{d}(\omega)$,
$\Phi_{3}(\omega)=\Phi_{2}(-\omega)$, and $\Phi_{4}(\omega)=f(\omega)f(-\omega)H_{d}(\omega)H_{d}(-\omega)$,
we get \eqref{eq:cov_avg_oupin_t}. $\C(\omega)$ for output noise
\eqref{eq:avg_cross_spectrum_on} is obtained by multiplying the expression
for $\C(\omega)$ for input noise with $H_{d}^{-1}(\omega)H_{d}^{-1}(-\omega)=(1+\omega^{2}\tau^{2})$.
In order to perform the back Fourier transformation one first needs
to rewrite the cross spectrum in order to isolate the frequency independent
term and the two terms that vanish for either $t<d$ or $t>d$, as
described in \nameref{sub:fourier_back_transformation},

\begin{align*}
\C(\omega)= & f(\omega)(\Em+Kw\left(\begin{array}{cc}
\gamma g & -\gamma g\\
1 & -1
\end{array}\right)H_{d}(\omega))\M\D\M^{T}f(-\omega)(\Em+Kw\left(\begin{array}{cc}
\gamma g & 1\\
-\gamma g & -1
\end{array}\right)H_{d}(-\omega))\\
+ & f(\omega)(\Em+Kw\left(\begin{array}{cc}
\gamma g & -\gamma g\\
1 & -1
\end{array}\right)H_{d}(\omega))\M\D\\
+ & \D\M^{T}f(-\omega)(\Em+Kw\left(\begin{array}{cc}
\gamma g & 1\\
-\gamma g & -1
\end{array}\right)H_{d}(-\omega))+\D\\
= & f(\omega)\M\D\M^{T}f(-\omega)+f(\omega)\M\D+\D\M^{T}f(-\omega)+\D,
\end{align*}
where in the last step we used $\left(\begin{array}{cc}
\gamma g & -\gamma g\\
1 & -1
\end{array}\right)\M=0$, because $\M$ is symmetric, obtaining \eqref{eq:cov_avg_oupon_t}.
For each of the first three terms in the last expression the right
integration contour needs to be chosen as described in \nameref{sub:fourier_back_transformation}
on the example of the general expression \eqref{eq:cross_spectrum_on_separated}.

\subsection{Implementation of noisy rate models\label{sub:implementation_noisy_rate}}

The dynamics is propagated in time steps of duration $\Delta t$
(note that in other works we use $h$ as a symbol for the computation
step size, which here is used as the symbol for the kernel). The product
of the connectivity matrix with the vector of output variables at
the end of the previous step $i-1$ is the vector $\I(t_{i})$ of
inputs at the current step $i$. The intrinsic time scale of the system
is determined by the time constant $\tau$. For sufficiently small
time steps $\Delta t\ll\tau$ these inputs can be assumed to be time
independent within one step. So we can use \eqref{eq:rdon} or \eqref{eq:rdin}
and analytically convolve the kernel function $h$ assuming the input
to be constant over the time interval $\Delta t$. This corresponds
to the method of exponential integration \citep[see App. C.6]{Rotter99a}
requiring only local knowledge of the connectivity matrix $\w$. Note
that this procedure becomes exact for $\Delta t\rightarrow0$ and
for finite $\Delta t$ is an approximation. The propagation of the
initial value $r_{j}(t_{i-1})$ until the end of the time interval
takes the form $r_{j}(t_{i-1})\, e^{-\Delta t/\tau}$ because $h(t_{i})=h(t_{i-1})\, e^{-\Delta t/\tau}$,
so we obtain the expression $r_{j}(t_{i})$ at the end of the step
as

\begin{equation}
r_{j}(t_{i})=e^{-\Delta t/\tau}\, r_{j}(t_{i-1})+(1-e^{-\Delta t/\tau})\, I_{j}(t_{i}),\label{eq:numeric_OUP}
\end{equation}
where $I_{j}$ denotes the input to the neuron $j$. For output noise
the output variable of neuron $j$ is $y_{j}=r_{j}+x_{j}$, with the
locally generated additive noise $x_{j}$ and hence the input is $I_{j}(t_{i})=(\w\,\y(t_{i}))_{j}$.
In the case of input noise the output variable is $r_{j}$ and the
additional noise is added to the input variable, $I_{j}(t_{i})=(\w\,\r(t_{i}))_{j}+x_{j}(t_{i})$.
In both cases $x_{j}$ is implemented as a binary noise: in each time
step, $x_{j}$ is independently and randomly chosen to be $1$ or
$-1$ with probability $0.5$ multiplied with $\rho/\sqrt{\Delta t}$
to satisfy \eqref{eq:noise} for discretized time. Here the $\delta$-function
is replaced by a {}``rectangle'' function that is constant on the
interval of length $\Delta t$, vanishes elsewhere, and has unit integral.
The factor $\Delta t^{-1}$ in the expression for $x^{2}$ ensures
the integral to be unity. So far, the implementation assumes the synaptic
delay to be zero. To implement a non-zero synaptic delay $d$, each
object representing a neuron contains an array $b$ of length $l_{d}=d/\Delta t$
acting as a ring buffer. The input $I_{j}(t_{i})$ used to calculate
the output rate at step $i$ according to \eqref{eq:numeric_OUP}
is then taken from position $i\ \mathrm{mod}\ l_{d}$ of this array
and after that replaced by the input presently received from the network,
so that the new input will be used only after one delay has passed.
This sequence of buffer handling can be represented as 
\begin{align*}
I_{j}(t_{i}) & \leftarrow b[i\,\mathrm{mod}\, l_{d}]\\
b[i\,\mathrm{mod}\, l_{d}] & \leftarrow\begin{cases}
(\w\,\r)_{j}+x_{j} & \quad\text{for input noise}\\
(\w\,\y)_{j} & \quad\text{for output noise}
\end{cases}.
\end{align*}
The model is implemented in Python version 2.7 \citep{Python} using
numpy 1.6.1 \citep{numpy} and scipy 0.9.0 \citep{scipy01}.

\subsection{Implementation of binary neurons in a spiking simulator code\label{sub:implementation_binary}}

The binary neuron model is implemented in the NEST simulator, version
2.2.1 \citep{Gewaltig_07_11204}, which allows distributed simulation
on parallel machines and handles synaptic delays in the established
framework for spiking neurons \citep{Morrison05a}. The name of the
model is {}``\texttt{ginzburg\_neuron}''. In NEST information is
transmitted in form of point events, which in case of binary neurons
are sent if the state of the neuron changes: one spike is sent for
a down-transition and two spikes at the same time for an up-transition,
so the multiplicity reflects the type of event. The logic to decode
the original transitions is implemented in the function $\mathrm{handle}$
shown in \prettyref{alg:handle_function}. If a single spike is received,
the synaptic weight $w$ is subtracted from the input buffer at the
position determined by the time point of the transition and the synaptic
delay. In distributed simulations a single spike with multiplicity
$2$ sent to another machine is handled on the receiving side as two
separate events with multiplicity $1$ each. In order to decode this
case on the receiving machine we memorize the time ($t_{\mathrm{last}}$)
and origin (global id $\mathrm{gid}_{\mathrm{last}}$ of the sending
neuron) of the last arrived spike. If both coincide to the spike under
consideration, the sending neuron has performed an up transition $0\rightarrow1$.
We hence add twice the synaptic weight $2w$ to the input buffer of
the target neuron, one that reflects the real change of the system
state and another that compensates the subtraction of $w$ after reception
of the first spike of a pair. The algorithm relies on the fact that
within NEST two spikes that are generated by one neuron at the same
time point are delivered sequentially to the target neurons. This
is assured, because neurons are updated one by one: The update propagates
each neuron by a time step equal to the minimal delay $d_{\mathrm{min}}$
in the network. All spikes generated within one update step are written
sequentially into the communication buffers, and finally the buffers
are shipped to the other processors \citep{Morrison05a}. Hence a
pair of spikes generated by one neuron within a single update step
will be delivered consecutively and will not be interspersed by spikes
from other neurons with the same time stamp.

The model exhibits stochastic transitions (at random points in time)
between two states. The transitions are governed by probabilities
$\phi(h)$. Using asynchronous update \citep{PDP86a}, in each infinitesimal
interval $[t,t+\delta t)$ each neuron in the network has the probability
$\frac{1}{\tau}\delta t$ to be chosen for update \citep{Hopfield82}.
A mathematically equivalent formulation draws the time points of update
independently for all neurons. For a particular neuron, the sequence
of update points has exponentially distributed intervals with mean
duration $\tau$, i.e. it forms a Poisson process with rate $\tau^{-1}$.
We employ the latter formulation to incorporate binary neuron models
in the globally time-driven spiking simulator NEST \citep{Gewaltig_07_11204}
and constrain the points of transition to a discrete time grid $\Delta t=0.1\ms$
covering the interval $d_{\mathrm{min}}\geq\Delta t$. This neuron
state update is implemented by the algorithm shown in \prettyref{alg:neuron_update}.
Note that the field $h$ is updated in steps of $\Delta t$ while
the activity state is updated only when the current time exceeds the
next potential transition point. As the last step of the activity
update we draw an exponentially distributed time interval to determine
the new potential transition time. The potential transition time is
represented with a higher resolution (on the order of microseconds)
than $\Delta t$ to avoid a systematic bias of the mean inter-update-interval.
This update scheme is identical to the one used in \citep{Hopfield82}.
Note that the implementation is different from the classical asynchronous
update scheme \citep{Vreeswijk98}, where in each discrete time step
$\Delta t$ exactly one neuron is picked at random. The mean inter-update-interval
(time constant $\tau$ in \prettyref{alg:neuron_update}) in the latter
scheme is determined by $\tau=\Delta tN$, with $N$ the number of
neurons in the network. For small time steps both schemes converge
so that update times follow a Poisson process.

At each update time point the neuron state becomes $1$ with the probability
given by the function $\phi$ applied to the input at that time according
to \eqref{eq:input_functional} and $0$ with probability $1-\phi$.
The input is a function of the whole system state and is constant
between spikes which indicate state changes. Each neuron therefore
maintains a state variable $h$ at each point in time holding the
summed input and being updated by adding and subtracting the input
read from the ring buffer $b$ at the point $\mathrm{readpos(t)}$
corresponding to the current time \citep[see ][for the implementation of the ring buffer, i.p. Fig 6]{Morrison05a}.
The ring buffer enables us to implement synaptic delays. For technical
reasons this implementation requires a minimal delay of a single simulation
time step \citep{Morrison08_267}. The gain function $\phi$ applied
to the input $h$ has the form 

\begin{equation}
\phi(h)=c_{1}h+c_{2}\frac{1}{2}\,(1+\tanh(c_{3}(h-\theta))),\label{eq:tanh}
\end{equation}
where throughout this manuscript we used $c_{1}=0$, $c_{2}=1$, and
$c_{3}=\beta$, as defined in \nameref{sub:parameters}.

\begin{algorithm}[h]
\begin{lstlisting}[language=Python,mathescape=true,numbers=left,tabsize=4]
$y \leftarrow 0$    // initially neuron is inactive
$t_\mathrm{next} \leftarrow - \tau \log( \mathrm{rand}() )$ // next time point of update

for each time step $t$:
	
   $h \leftarrow h + b [ \mathrm{readpos}(t) ]$

   if $t > t_\mathrm{next}$:
		// up-state with probability given by
		// gain function $\phi$ depending on input $h(t)$
        
		if $\phi(h) > \mathrm{rand}()$:
			$y_\mathrm{new} \leftarrow 1$
		else:  
			$y_\mathrm{new} \leftarrow 0$

		if $y_\mathrm{new} \neq y$:
			// down transition: send single spike
			// up transition: send two spikes 
			send ($y_\mathrm{new} + 1$ spikes)

			$y \leftarrow y_\mathrm{new}$

		// add an exponentially distributed time interval 
		$t_\mathrm{next} \leftarrow t_\mathrm{next} - \tau \log ( \mathrm{rand}() )$
\end{lstlisting}

\caption{Update function of a binary neuron embedded in the spiking network
simulator NEST. The function $\mathrm{readpos(t)}$ returns a position
in the ring buffer $b$ corresponding to the current time point.\label{alg:neuron_update}}
\end{algorithm}

\begin{algorithm}[h]
\begin{lstlisting}[language=Python,mathescape=true,numbers=left,tabsize=4]
handle($t_\mathrm{spike}, d, \mathrm{gid}, m$):

	if $m = 1$:

	// multiplicity = 1, either a single $1 \rightarrow 0$ event 
	// or the first or second of a pair of $0 \rightarrow 1$ events

		if $\mathrm{gid} = \mathrm{gid}_\mathrm{lastspike}$ and $t_\mathrm{spike} = t_\mathrm{lastspike}$:

			// received twice the same event, so transition $0 \rightarrow 1$
			// add $2 w$ to compensate for subtraction after reception
			// of first event
	    	$b [ \mathrm{pos}(t_\mathrm{spike}, d, t) ] \leftarrow b[ \mathrm{pos}(t_\mathrm{spike}, d, t) ] + 2 w$

		else:
			// count this event negatively,
			// assuming it comes as single event
	    	// transition $1 \rightarrow 0$
			$b [ \mathrm{pos}(t_\mathrm{spike}, d, t) ] \leftarrow b[ pos(t_\mathrm{spike}, d, t) ] - w$
	else: 
	
	// multiplicity != 1
	
		if m = 2:

			// count this event positively, transition $0 \rightarrow 1$
			$b [ pos(t_\mathrm{spike}, d, t) ] \leftarrow b [ pos(t_\mathrm{spike}, d, t) ] + w$
			$\mathrm{gid}_\mathrm{lastspike} \leftarrow \mathrm{gid}$
			$t_\mathrm{lastspike} \leftarrow t_\mathrm{spike}$
\end{lstlisting}

\caption{Input spike handler of a binary neuron embedded in the spiking network
simulator NEST. The simulation kernel calls the handle function for
each spike event to be delivered to the neuron. A spike event is characterized
by the time point of occurrence $t_{\mathrm{spike}}$, the synaptic
delay $d$ after which the event should reach the target, the global
id $\mathrm{gid}$ identifying the sending neuron, and the multiplicity
$m\ge1$, indicating the reception of multiple spike events. The function
$\mathrm{pos}(t_{\mathrm{spike}},d,t)$ returns the position in the
ring buffer $b$ to which the spike is added so that it will be read
at time $t+d$ by the update function of the neuron, see \prettyref{alg:neuron_update}.\label{alg:handle_function}}
\end{algorithm}

\subsection{Implementation of Hawkes neurons in a spiking simulator code\label{sub:implementation_hawkes}}

Hawkes neurons \citep{Hawkes71_438} were introduced in the NEST simulator
in version 2.2.0 \citep{Gewaltig_07_11204}. The name of the model
is {}``\texttt{pp\_psc\_delta}''. In the following we describe the
implemented neuron model in general and mention the particular choices
of parameter and correspondences to the theory presented in \nameref{sec:equivalence_hawkes_oup}.
The dynamics of the quasi-membrane potential $u$ is integrated exactly
within a time step $\Delta t$ of the simulation \citep{Rotter99a},
expressing the voltage $u(t_{i})$ at the end of time step $i$ by
the membrane potential at the end of the previous time step $u(t_{i-1})$
as

\begin{equation}
u(t_{i})=e^{-\Delta t/\tau}\, u(t_{i-1})+(1-e^{-\Delta t/\tau})\, R_{m}I_{e}+b(t_{i}),
\end{equation}
where $I_{e}$ is a time-step wise constant input current (equal to
$0$ in all simulations presented in this article) and $R_{m}=\taum/C_{m}$
is the membrane resistance. The buffer $b(t_{i})$ contains the summed
contributions of incoming spikes, multiplied by their respective synaptic
weight, which have arrived at the neuron within the interval $(t_{i-1},t_{i}]$.
$b$ is implemented as a ring-buffer in order to handle the synaptic
delay, logically similar as in \nameref{sub:implementation_noisy_rate},
described in detail in \citet{Morrison05a}. The instantaneous spike
emission rate is $\lambda=[c_{1}u+c_{2}e^{c_{3}u}]_{+}$, where we
use $c_{3}=0$ in all simulations presented here. The quantities in
the theory \nameref{sec:equivalence_hawkes_oup}, in particular in
\eqref{eq:hawkes_def}, are related to the parameters of the simulated
model in the following way. The quantity $r$ relates to the membrane
potential $u$ as $r=c_{1}u+c_{2}$ and the background rate $\nu$
agrees to $c_{2}=\nu$. Hence the synaptic weight $J_{ij}$ corresponds
to the synaptic weight in the simulation multiplied by $c_{1}$. For
the correspondence of the Hawkes model to the OUP with output noise
of variance $\rho^{2}$ we use \eqref{eq:prob_positive_rate_Hawkes}
to adjust the background rate $\nu$ in order to obtain the desired
rate $\lambda_{0}=\rho^{2}$ and we choose the synaptic weight $J$
of the Hawkes model so that the linear coupling strength $w$ of the
OUP agrees to the effective linear weight given by \eqref{eq:hawkes_lin_weight}.
These two constraints can be fulfilled simultaneously by solving \eqref{eq:prob_positive_rate_Hawkes}
and \eqref{eq:hawkes_lin_weight} by numerical iteration. The spike
emission of the model is realized either with or without dead time.
In this article we only used the latter. In the presence of a dead
time, which is constrained to be larger than the simulation time step,
at most one spike can be generated within a time step. A spike is
hence emitted with the probability $p_{\ge1}=1-e^{\lambda\Delta t}$,
where $e^{\lambda\Delta t}$ is the probability of the complementary
event (emitting $0$ spikes), implemented by comparing a uniformly
distributed random number to $p_{\ge1}$. The refractory period is
handled as described in \citet{Morrison05a}. Without refractoriness,
the number of emitted spikes is drawn from a Poisson distribution
with parameter $\lambda\Delta t$, implemented in the GNU Scientific
Library \citep{GSL06}. Reproducibility of the random sequences for
different numbers of processes and threads is ensured by the concept
of random number generators assigned to virtual processes, as described
in \citep{Plesser07_672}.

\subsection{Parameters of simulations\label{sub:parameters}}

For all simulations we used $\gamma=0.25$ corresponding to the biologically
realistic fraction of inhibitory neurons, a connectivity probability
$p=0.1$, and a simulation time step of $\Delta t=0.1\ms$. For binary
neurons we measured the covariance functions with a resolution of
$1\ms$, for all other models the resolution is $0.1\ms$. Simulation
time is $10,000\ms$ for linear rate and for LIF neurons, $50,000\ms$
for Hawkes, and $100,000\ms$ for binary neurons. The covariance is
obtained for a time window of $\pm100\ms$.

The parameters for simulations of the LIF model presented in \prettyref{fig:covariance_spiking}
and \prettyref{fig:echoes} are $J=0.1\mV,$ $\tau=20\ms,$ $\tau_{s}=2\ms,$
$\tau_{r}=2\ms,$ $V_{\theta}=15\mV,$ $V_{r}=0,$ $g=6,$ $d=3\ms,$
$N=8000$. The number of neurons in the corresponding networks of
other models is the same. Cross covariances are measured between the
summed spike trains of two disjoint populations of $N_{\mathrm{rec}}=1000$
neurons each. The single neuron autocovariances $a_{\alpha}$ are
averaged over a subpopulation of $100$ neurons. The autocovariances
of the population averaged activity $\frac{1}{N_{\alpha}}a_{\alpha}+C_{\alpha\alpha}$
for population $\alpha\in\{\Ex,\In\}$ (shown in \prettyref{fig:covariance_spiking})
are constructed from the estimated single neuron population averaged
autocovariances $a_{\alpha}$ and cross covariances $C_{\alpha\alpha}$.
This enables us to estimate $a_{\alpha}$ and $C_{\alpha\alpha}$
from the activity of a small subpopulation and still assigns the correct
relative weights to both contributions. The corresponding effective
parameters describing the system dynamics are $\mu=15\mV,$ $\sigma=10\mV,$
$r=23.6\Hz$ (see \prettyref{eq:diffeq_iaf} and the following text
for details). 

The parameters of the Hawkes model and of the noisy rate model with
output noise yielding quantitatively agreeing covariance functions
are:
\begin{itemize}
\item For simulations of the noisy rate model with output noise presented
in \prettyref{fig:covariance_spiking} and \prettyref{fig:power_spectra}
the parameters are $w=0.0043,$ $g\thickapprox5.93,$ $\tau=4.07\ms,$
$\rho^{2}=23.6\Hz,$ $d=3\ms$ (see \prettyref{eq:rdon}, \prettyref{eq:rdin}).
In \prettyref{fig:power_spectra} also results for $d=1\ms$ and for
input noise are shown. Signals are measured from $N_{\mathrm{rec}}=500$
neurons in each population to obtain $c_{\Ex\In},\ c_{\In\Ex}$ and
from the whole population to determine $c_{\Ex\Ex}$ and $c_{\In\In}$.
The cross covariances $C_{\Ex\Ex}$ and $C_{\In\In}$ are estimated
from two disjoint subpopulations each comprising half of the neurons
of the respective population.
\item For the network of Hawkes neurons presented in \prettyref{fig:covariance_spiking}
we used $\lambda_{0}\thickapprox22.54\Hz$ (see \prettyref{eq:prob_positive_rate_Hawkes}),
$J=0.0055\mV$, $d=3\ms$, and the same $g$ and $\tau$ as for the
noisy rate model. We measured the cross covariances in the same way
as for the LIF model, but using the spike trains from sub-populations
of $N_{\mathrm{rec}}=2000$ neurons. The autocovariances of the population
averaged activity were estimated from the whole populations.
\end{itemize}
The network of binary neurons shown in \prettyref{fig:echoes} uses
$\theta=-3.89\mV,$ $\beta=0.5\mV{}^{-1},$ $J=0.02\mV,$ $d=3\ms$
(see \prettyref{eq:input_functional}, \prettyref{eq:tanh}), and
the same $g$ and $\tau$ as the noisy rate model. Covariances are
measured using the signals from all neurons.

The simulation results for the network of binary neurons presented
in \prettyref{fig:binandlin} uses $\theta=-2.5\mV,$ $\tau=10\ms,$
$\beta=0.5\mV{}^{-1},$ $g=6,$ $J\thickapprox0.0447\mV,$ $N=2000$
and the smallest possible value of synaptic delay is $d=0.1\ms$ equal
to time resolution (the same set of parameters only with modified
$\beta=1\mV^{-1}$ was used to create \prettyref{fig:linofbin}).
The cross covariances $C_{\Ex\Ex}$ and $C_{\In\In}$ are estimated
from two disjoint subpopulations each comprising half of the neurons
of the respective population, $c_{\Ex\In}$ is measured between two
such subpopulations. For $c_{\Ex\Ex}$ and $c_{\In\In}$ we used the
full populations.

The parameters required for a quantitative agreement with the rate
model with input noise are $w\thickapprox0.011,$ $\rho\thickapprox2.23\,\mathrm{\sqrt{ms}}$.
We used the same parameters in \prettyref{fig:invsoutdegree}, where
additionally results for $w=0.018$ are shown. The population sizes
are the same as for the binary network. The covariances are estimated
in the same way as for the rate model with output noise. Note that
the definition of noisy rate models has no limitation for units of
$\rho^{2}$. These can be arbitrary and are chosen differently as
required by the correspondence with either spiking or binary neurons.

\section*{Acknowledgments\pdfbookmark[1]{Acknowledgements}{AcknowledgementsPage}}

We gratefully appreciate ongoing technical support by our colleagues
in the NEST Initiative, especially Moritz Deger for the implementation
of the Hawkes model. Binary and spiking network simulations performed
with NEST (www.nest-initiative.org). Partially supported by the Helmholtz
Association: HASB and portfolio theme SMHB, the Jülich Aachen Research
Alliance (JARA), the Next-Generation Supercomputer Project of MEXT,
and EU Grant 269921 (BrainScaleS).

\clearpage{}

\pdfbookmark[1]{References}{ReferencesPage} 
\end{document}